\documentclass[letterpaper,10pt]{article} 

\usepackage{opticameet3} 
\usepackage{braket}
\usepackage{floatpag}

\usepackage{ccaption}

\usepackage[font=footnotesize,labelfont=bf]{caption}

\newcommand\authormark[1]{\textsuperscript{#1}}

\usepackage{amsmath,amssymb}
\usepackage[colorlinks=true,bookmarks=false,citecolor=blue,urlcolor=blue]{hyperref} 

\begin{document}

\title{3D-Patterned Inverse-Designed Mid-Infrared Metaoptics}

\author{Gregory Roberts,\authormark{1} Conner Ballew,\authormark{1,2} Tianzhe Zheng,\authormark{1} Juan C. Garcia,\authormark{3} Sarah Camayd-Muñoz,\authormark{1,4} Philip W. C. Hon,\authormark{3} and Andrei Faraon\authormark{1,*}}

\address{\authormark{1}Kavli Nanoscience Institute and Thomas J. Watson Sr. Laboratory of Applied Physics, California Institute of Technology, 1200 E California Blvd, Pasadena, CA 91125\\
\authormark{2}Currently with Jet Propulsion Laboratory, California Institute of Technology, 4800 Oak Grove Dr, Pasadena, CA 91109\\
\authormark{3}NG Next, Northrop Grumman Corporation, 1 Space Park Drive, Redondo Beach, CA 90278\\
\authormark{4}Currently with Applied Physics Laboratory, The Johns Hopkins University, 11100 Johns Hopkins Road, Laurel, MD 20723}

\email{\authormark{*}faraon@caltech.edu}

\begin{abstract}
Modern imaging systems can be enhanced in efficiency, compactness, and application through introduction of multilayer nanopatterned structures for manipulation of light based on its fundamental properties.  High transmission efficiency multispectral imaging is surprisingly elusive due to the commonplace use of filter arrays which discard most of the incident light.  Further, most cameras do not leverage the wealth of information in polarization and spatial degrees of freedom.  Optical metamaterials can respond to these electromagnetic properties but have been explored primarily in single-layer geometries, limiting their performance and multifunctional capacity.  Here we use advanced two-photon lithography to realize multilayer scattering structures that achieve highly nontrivial optical transformations intended to process light just before it reaches a focal plane array.  Computationally optimized multispectral and polarimetric sorting devices are fabricated with submicron feature sizes and experimentally validated in the mid-infrared.  A final structure shown in simulation redirects light based on its angular momentum.  These devices demonstrate that with precise 3-dimensional nanopatterning, one can directly modify the scattering properties of a sensor array to create advanced imaging systems.
\end{abstract}

\section*{Main}
Nanophotonics synthesizes the study of light-matter interaction with the precise, repeatable techniques of nanofabrication.  For example, dielectric metasurfaces are arrays of subwavelength scatterers that apply a spatially varying phase, polarization or amplitude response to an incoming wavefront \cite{RecentMetaReview}.  The local control is related to the specific shape of each scatterer which can be chosen to compactly replicate and combine functionalities of common optical components like lenses, beamsplitters, polarizers, and waveplates or realize more novel devices such as those used for visible color routing at the pixel level \cite{NTTRecentRouter}.  For metasurfaces, the absence of substantial inter-element electromagnetic coupling is often leveraged for ease of design, but this simplification also limits the available degrees of freedom.  Ultimately, we would like to tailor unique scattering behaviors for wavefronts with different spectral, spatial, and polarization properties.  To do this, we can expand the design space to volumetric devices where a material is patterned at subwavelength resolution in three dimensions.

Three-dimensional (3D) devices take advantage of a larger set of optical modes to achieve unprecedented performance in a variety of tasks, but require an efficient gradient-based optimization algorithm based on full-wave electromagnetic simulation.  Searching the high-dimensional space of permittivity profiles, typically for a local optimum to an electromagnetic merit function, is referred to as inverse design \cite{MillerThesis,MoleskyInvDesReview}.  In this area, quasi-2D on-chip photonic devices have been explored extensively where patterning in the direction of light propagation is achieved in a single fabrication layer \cite{VuckovicSpectralDemultiplex,YablanovitchInvDesFoundation}.  The fully 3D design paradigm for free space applications is yet to emerge, mostly due to the increased fabrication complexity of volumetric devices.  However, early works in this area utilizing one- and two-layer processes for optical applications or many-layer microwave prototypes have shown the utility of moving to thicker devices \cite{SellFanGrating,Arbabi2LayerMeta,FaraonOpticalVolumetric}.  In this work, we optimized a two-photon polymerization (TPP) lithography process to create intricate, multilayer structures at optical wavelengths.  This technique has been employed in the past for fabricating refractive, diffractive, gradient index, and extruded 2D inverse-designed optical components \cite{NanoscribeStackedRefractive,NanoscribeStackedDiffractive,BraunScribe,2DConcentrator}.  
By exploiting TPP flexibility for 3D patterning at subwavelength resolution, we experimentally demonstrated multiple inverse-designed, multilayer photonic devices with applications to advanced imaging in the mid-infrared band (3-6$\mu m$).

Compact imaging systems typically contain wavelength- and polarization-selective elements to characterize fundamental properties of wavefronts. Color imaging in consumer cameras follows this approach where absorptive filters are placed on top of collections of pixels to sense three or four spectral overlaps.  The classic arrangement, referred to as a Bayer pattern, consists of a ‘red’, ‘blue’, and two ‘green’ filters arranged 2x2 in a square \cite{BayerPatent}.  Filtering schemes like this come at a cost of transmission efficiency because they absorb all light outside of their passband leading to average transmission values of approximately 33$\%$ under uniform spectral illumination.  Solutions to this problem have converged on the concept of color routing where scattering structures accept light incident on a group of pixels and redirect each wavelength band to a different pixel \cite{PanasonicRouter,NTTRecentRouter,FanFreeformSplitters,ColorSplitterSims,FaraonOpticalVolumetric}.  In this manuscript, we demonstrate an efficient, multilayer inverse-designed device in the mid-infrared for accomplishing this task and further augment it to sense linear polarization. Beyond multispectral imaging, the geometry of splitting light at the the focal plane, depicted in \verb+Fig. 1a+, can be tailored to efficiently decode other electromagnetic properties.  Designing at the pixel level modularizes the optical system, allowing focal plane arrays equipped with arrangements of scattering structures to control the imaging properties of the camera.  \verb+Fig. 1b-d+ indicates the breadth of devices in this manuscript.

The first application we explored is combined multispectral and polarization imaging.  Absorption spectra in the mid-infrared, part of the molecular fingerprinting region \cite{MIRFingerprintRef}, correlate strongly to distinct chemical species.  Among many areas of interest, this can be used for environmental monitoring \cite{DetectCO2,MetamaterialAbsorbGas} and biomedical imaging \cite{CancerDetectionMIR,FTIRBioMaterials}.  Solutions such as multiplexed filters in the mid-infrared suffer from low overall transmission efficiency \cite{MIRPlasmonic}.  They also lack a straightforward path towards multifunctionality that may be critical for a given application.  In remote thermal monitoring, for example, multispectral and polarization filtering can be used in tandem to distinguish radiated and reflected light reducing instances of thermal blindness \cite{ThermalBlindness}. To address these challenges, we designed and fabricated a multilayer color-routing device with additional linear polarization discrimination.

The optimization goal, stated in \verb+Eq. 1+, is constructed to sort three spectral bands from $3.7-5\mu m$ and distinguish between linear polarization for the middle band.  The device dimensions are $30.15 \mu m$ x $30.15 \mu m$ x $18 \mu m$ ($6.6$ x $6.6$ x $4.0$ $\lambda_{mid}^3$), broken into six $3 \mu m$ thick layers, compact enough to be tiled on a high resolution focal plane array.

\begin{equation}
\begin{aligned}
    \max_{\Vec{\epsilon} \in \{\epsilon_{min}, \epsilon_{max}\}^N} g(\Vec{E}) = \sum_{\lambda} S((\sum_p \sum_q \kappa(q, p,\lambda) \frac{I_p(\Vec{r_q}, \lambda)}{I_{max}(\lambda)}); k) \\
    I_p(\Vec{r_q}, \lambda) = ||\Vec{E_p}(\Vec{r_q}, \lambda)||^2
\end{aligned}
\end{equation}

\noindent{}The adjoint method for electromagnetics, aims to efficiently optimize merit functions like this, where device performance is phrased in terms of electric and magnetic fields in an observation region \cite{MillerThesis}.  Here, the electric field intensity at the center of each quadrant is maximized for correct wavelengths and polarizations, and minimized for incorrect ones through choice of sign in the $\kappa(q,p,\lambda)$ weighting function where $p$ indexes the linear polarization and $q$ indexes the quadrant with center $\Vec{r_q}$.  The first summation targets broadband performance by including closely spaced wavelengths in each band to effectively optimize the device across a continuum.  The purpose of the softplus function, $S$, is described in the supplemental alongside other optimization figures of merit for this work \cite{SupplementaryReference}.  This optimization function is nonlinear over the high-dimensional (\textasciitilde  $10^4$-dimensional for devices in this work) permittivity tensor, composed of deeply subwavelength volumetric units (voxels).  It is optimized via gradient descent enabled by the well-known adjoint method \cite{MillerThesis,YablanovitchInvDesFoundation,VuckovicSpectralDemultiplex}.  Combining the electric fields in the device from adjoint simulations with those from the expected illumination, in this case broadband linearly polarized plane waves, the gradient is computed in a fixed number of simulations independent of the number of design voxels.  Fabrication constraints were incorporated for layering, feature size control, and binarization using averaging, lateral maximum blurring, and sigmoid projection filters, respectively \cite{ProjectionMethods}.

The optimization results are shown in \verb+Fig. 2a,b+, where three sorting bands are present with the middle band focal spot conditioned on linear polarization.  Following this result, the device was fabricated using the Nanoscribe Photonic Professional GT, where subwavelength features in the mid-infrared are readily created in a proprietary IP-Dip polymer with low loss from roughly 3.5-5.5$\mu m$ \cite{IPDipIndex}.  Using a photolithography-based liftoff procedure, a series of $30\mu m$ diameter circular aluminum apertures were fabricated on a sapphire substrate.  Apertures, also included in the optimization, restrict the illumination to single devices for imaging and experimental power calibration.  The Nanoscribe was aligned to write devices directly on top of the apertures.  \verb+Fig. 2c+ shows scanning electron microscope (SEM) images of fabricated devices.  Each design was illuminated by a quantum cascade laser (QCL) with a beam waist on the order of the device size defocused such that the apertures were overfilled and sampled a roughly flat amplitude and phase section of the diverging beam.  This is intended to mimic the plane wave input used for device optimization.  The QCL can be tuned spectrally to probe the device at different wavelengths and the addition of linear polarizers and waveplates were used to control the input polarization.  Various focal planes of the device were imaged by a zinc selenide (ZnSe) aspheric lens onto a focal plane array (see \verb+Fig. S1+).

\verb+Fig. 2d,e+ contains the experimental spectral and polarization sorting efficiency, overall focal transmission, and focal spot intensities to compare to simulation.  Sorting efficiency measures the ratio of total focal plane signal incident on a given quadrant.  The experimental device matches the simulation well, but with reduced contrast.  This is likely due to imaging aberrations, experimental beam non-idealities and fabrication errors, which include device shrinkage and feature size mismatch from proximity effects and resolution limits \cite{TPPAccuracy}.  Transmission is measured as power through the device printed on top of a $30\mu m$ aperture that reaches the focal plane versus power through an empty $30\mu m$ aperture.  We attribute the uneven nature of the transmission in the experiment around $4.4 \mu m$ to minor power fluctuations or beam shifts between the device and pinhole normalization measurements.  In the focal plane images, one can see the focused spot move between the quadrants as the wavelength changes demonstrating the splitting functionality with the middle band sorted to opposite corners depending on its linear polarization.

For the second application, we investigated full Stokes imaging polarimetry, where one characterizes not only the linear polarization amplitudes, but also the phase relationship between them and the degree of polarization.  This rich information is widely applicable, including in areas of biomedical imaging and diagnosis \cite{CPCancerDiagnosis}, depth-based and facial recognition imaging \cite{PolarimetryFacialRec,DepthSensingPolCues}, atmospheric monitoring \cite{SkyMonitoringPol}, and bio-inspired polarization based navigation \cite{BioPolNavigation}.  In polarimetric imaging, the input state is cast in terms of a 4-dimensional vector containing its Stokes parameters, which together specify the orientation, handedness, and degree of polarization.  Complete reconstruction of this state is done through at least four independent measurements. Measurements can be multiplexed in time using a rotating waveplate \cite{RotatingStokes} or in space by dividing up the area on one or more focal plane arrays \cite{PolReviewPassive}.  The analogous geometry to using absorptive filters for color imaging is the division of focal plane (DoFP) technique where pixels are grouped together with each responsible for analyzing a specific polarization component.  Many implementations use micropolarizer elements as filters \cite{MicropolFiltering}, thus limiting the transmission efficiency of the camera to $50\%$ by rejecting orthogonally polarized light to each filter.  Lost transmission can be recovered using pixel-sized metasurface lenses that apply different phase masks to two orthogonal polarizations.  For example, six projections done pairwise onto orthogonal polarization basis states directly measure the four Stokes parameters \cite{EhsanPolarimetry}.  However, these six measurements contain redundant information which reduces camera resolution or degrades signal-to-noise ratio compared to a four-measurement device with the same overall size.  Recently, it was shown that a metasurface grating could project incident light onto four equally spaced analyzer states with each projection belonging to a different order \cite{PolCameraCapasso}.  This approach requires propagation to spatially separate each order and is inherently chromatic due to grating dispersion.  We adopted benefits and addressed shortcomings of both approaches by employing the modularity of a pixel-level design for adaptation of any camera sensor to full polarimetric imaging and utilizing a minimal four-state projection for maximal compactness.  Using only four measurements is, in the setting of highly optimized modern camera technology, a massive $33\%$ improvement in required chip area, or alternatively a large resolution or signal-to-noise ratio enhancement.  As an added benefit, inverse design provides a path towards broadband polarimetry, which is difficult to achieve with metasurface and waveplate based systems due to their inherent chromatic dispersion.

We optimized a device of size $30 \mu m$ x $30 \mu m$ x $18 \mu m$ in six $3 \mu m$ layers for this purpose, with the optimization figure of merit adapted to focus four analyzer polarization states to different quadrants and reject their orthogonal states to those same quadrants.  Further, we augmented the experimental system to probe arbitrary polarization states for different wavelengths depicted in \verb+Fig. S1, S2+.  The simulation and experimental results are presented in \verb+Fig. 3a-d+ and fabricated devices are shown in \verb+Fig. 3e+.  Performance is quantified with two metrics.  First, for each quadrant, the contrast, $C \in [-1, 1]$, is the transmission for an analyzer state versus its orthogonal state: $C = \frac{ T_{analyzer} - T_{orthogonal}}{T_{analyzer} + T_{orthogonal}}$.  The optimization solution performs better for the three elliptical polarizations compared to the circular polarization state in this case, likely due to a lack of degrees of freedom.  In supplementary \verb+Fig. S3+, we show a thicker 12-layer device with improved contrast of the circular polarization state.  Similar to the multispectral device, there is a reduced contrast experimentally which we attribute again to fabrication and experimental imperfections.  Second, transmission is quantified for each state, which, as shown in the supplement, is limited to $50\%$ in a perfect device due to required vector overlaps between analyzer states.  We note that this is not a limit on total device transmission, but simply a requirement of linearity.  Observing the focal plane images in \verb+Fig. 3c,d+ demonstrates the intuitive behavior of the device.  The most telling indication of desired behavior is seen by observing the orthogonal state inputs where the device can theoretically fully extinguish transmission to a quadrant \cite{SupplementaryReference}.  By comparison to the analyzer state, the same quadrant under each orthogonal state is convincingly dark. 
 Practical usage of and a calibration procedure for the device is demonstrated in simulation in the supplementary and analysis of reconstruction accuracy of pure and mixed polarization states is shown in \verb+Fig. S8+ and \verb+Fig. S9+, respectively \cite{SupplementaryReference}.

A third device that we explored only in simulation sorts on the spatial degree of freedom.  One property of wavefronts with spatial structure is their orbital angular momentum (OAM).  Beams with discrete OAM values are modeled as Laguerre-Gaussian (LG) modes, which comprise a set of orthogonal spatial modes in the paraxial wave equation \cite{LGModes,SupplementaryReference}.  These modes are candidates for free space optical communication networks where information can be multiplexed on both the OAM and spin degrees of freedom \cite{OAMCommReview}.  Isolated devices that efficiently demultiplex different angular momentum values in free space \cite{AMCommNatPhot} or from fibers \cite{OAMMuxDemuxFibers} are essential to high bandwidth communication links.  Further, efficient sorting devices have utility as multiplexing agents when used reciprocally with spatial light patterns projected into or emitted directly from their focal planes.  Additional communication bandwidth is achievable with further spatial multiplexing of angular momentum beamlets \cite{ScienceAMMultiplex}, where the receiver requires an array of devices with similar geometry to those shown in this manuscript.  Applicable to either the isolated or array geometry, we consider a routing structure sensitive to combinations of four OAM states and two spins in the form of circular polarization handedness.  \verb+Fig. 4a+ illustrates the optimized angular momentum sorting device, consisting of 8 design layers, each 2.4$\mu m$ thick and a $30.15\mu m$ x $30.15\mu m$ lateral aperture.  High sorting contrast is observed in \verb+Fig. 4b+, defined as transmission into the desired quadrant versus elsewhere in the focal plane for source $S_k$ and target quadrant $Q_k$, specifically $C_k=\frac{T_{Q_k} - \sum_{i\neq k}T_{Q_i}}{T_{Q_k} + \sum_{i\neq k}T_{Q_i}}$.  Each combination of OAM and spin is efficiently focused to a different quadrant as seen in \verb+Fig. 4c-f+.  Transmission values for a beamsplitting and subsequent filtering scheme as opposed to routing would be limited to 25$\%$ for each state, so the proposed device roughly doubles the signal-to-noise ratio of detection.  The response of the device to excitations with OAM and spin values different from the design points is analyzed in \verb+Fig. S11-12+.

In this work, we demonstrate multilayer, inverse-designed nanophotonic structures capable of augmenting both the performance and multifunctionality of imaging systems.  Using the same configuration of lenses inside a typical camera and replacing the scattering element on top of the focal plane array, this technology enables cameras sensitive to angular momentum, polarization state, arbitrary spectral signatures, or combinations of multiple electromagnetic properties.  There are exciting avenues for exploration in the mid-infrared where fabrication is accessible via TPP tools such as the Nanoscribe.  We envision targeting specific narrow absorption bands for applications in chemical and biomedical imaging and tiling different types of splitting elements in the same array.  Moving forward, we can think of these elements as part of a computational imaging system where we design efficient reconstruction problems by utilizing direct control over the scattering properties of an array of elements in the optical path \cite{ComputationalMetasurfaceInvDes}.

By utilizing a well-optimized fabrication procedure and additional design rules, TPP fabrication can be pushed to the near-infrared range \cite{2DConcentrator}.  Scaling to longer wavelengths in the infrared requires a polymer transparent beyond $5.5\mu m$ or the use of a material inversion technique \cite{PolymerInversion} and parallel writing strategies for feasible fabrication times \cite{ParallelTPP}.  Currently, necessity of industry-level fabrication procedures are a barrier to demonstrating volumetric inverse design for visible wavelengths.  Typical integrated circuits, like those found in modern computer processors, consist of greater than ten layers of precisely aligned subwavelength structures with respect to visible wavelengths \cite{CMOSPlasmonics}.  By replacing metals with transparent optical materials in silicon-based CMOS processes, these fabrication techniques can realize the types of structures shown in this work at an industrial scale.  Currently, cost, complexity, and availability of these fabrication methods limit the exploration of multilayer photonic devices in academic and prototype settings.  Advances in accessible multilayer fabrication is a worthwhile endeavor to unlock a broad spectrum of imaging applications \cite{SubtractivePhotonics}.  Beyond replacing traditional absorptive-based Bayer filters with color routing structures, optimized devices targeting structural color will impact reflective display technology \cite{StructuralColorReview} and efficient, spectrally-selective waveguide couplers will improve performance of augmented reality displays \cite{ARChallenges}.  We believe there is large, untapped potential for 3D, inverse-designed photonics in both research and commercial settings.  The present work is a substantial step towards the realization of these complex devices for real-world applications.


\begin{figure}
\thisfloatpagestyle{plain}
\begin{center}
    \includegraphics{"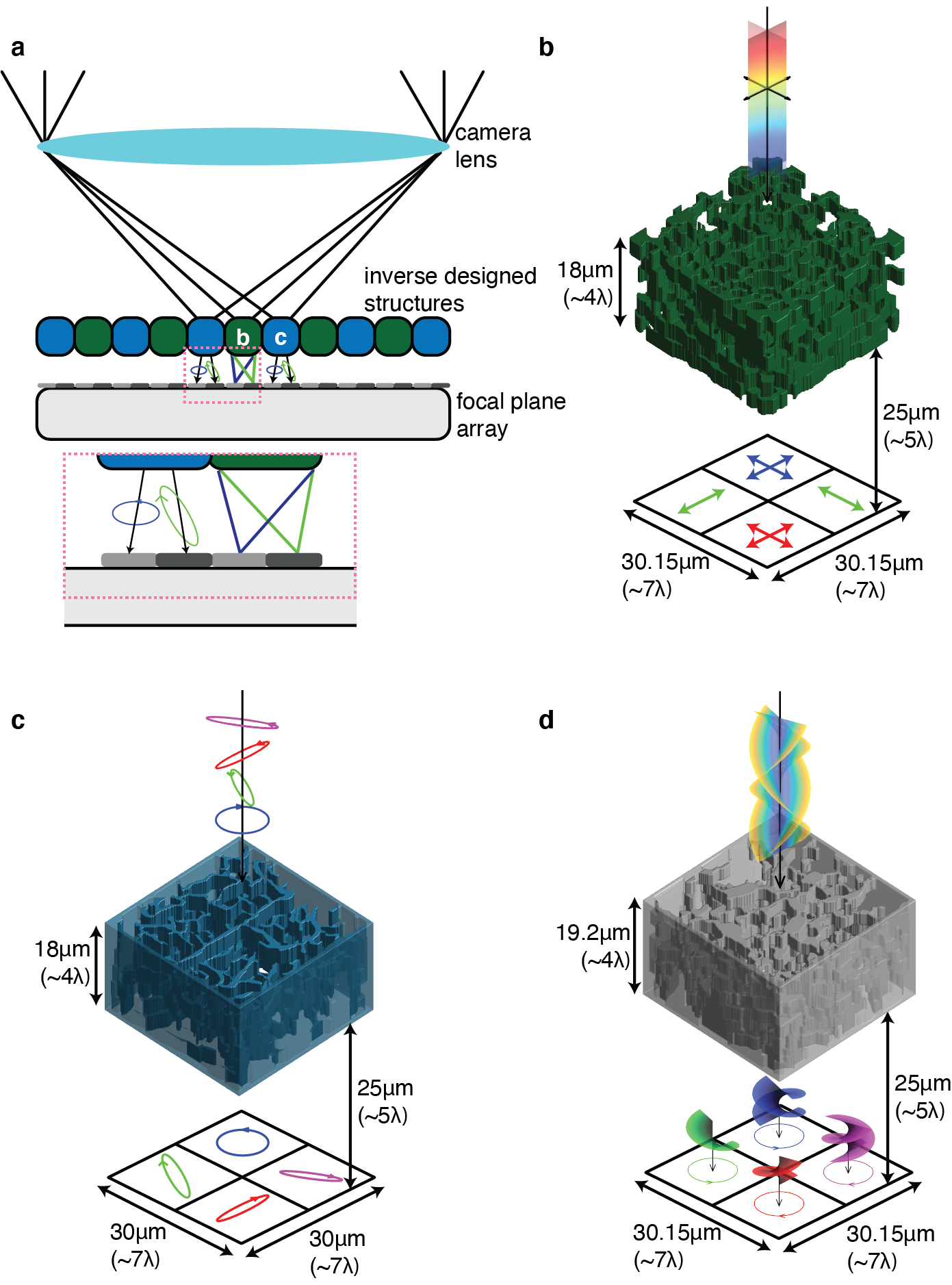"}
\end{center}
\caption{Conceptual depiction of devices in this work. \textbf{(a)} 2D cross section schematic of camera with inverse designed scattering elements placed on top of photosensitive elements at the focal plane of the imaging lens.  Green elements sort by color and blue elements sort by polarization, shown in more detail in (b, c).  \textbf{(b)} Rendering of multispectral and linear polarization device that sorts three bands of wavelengths with the middle band further split on polarization. \textbf{(c)} Rendering of full Stokes polarimetry device that sorts four analyzer Jones vectors to different quadrants. \textbf{(d)} Rendering of angular momentum splitting device that sorts combinations of orbital angular momentum ($l$) and spin ($s$) degrees of freedom.}
\end{figure}\label{fig1}

\begin{figure}
\thisfloatpagestyle{plain}
\begin{center}
    \includegraphics{"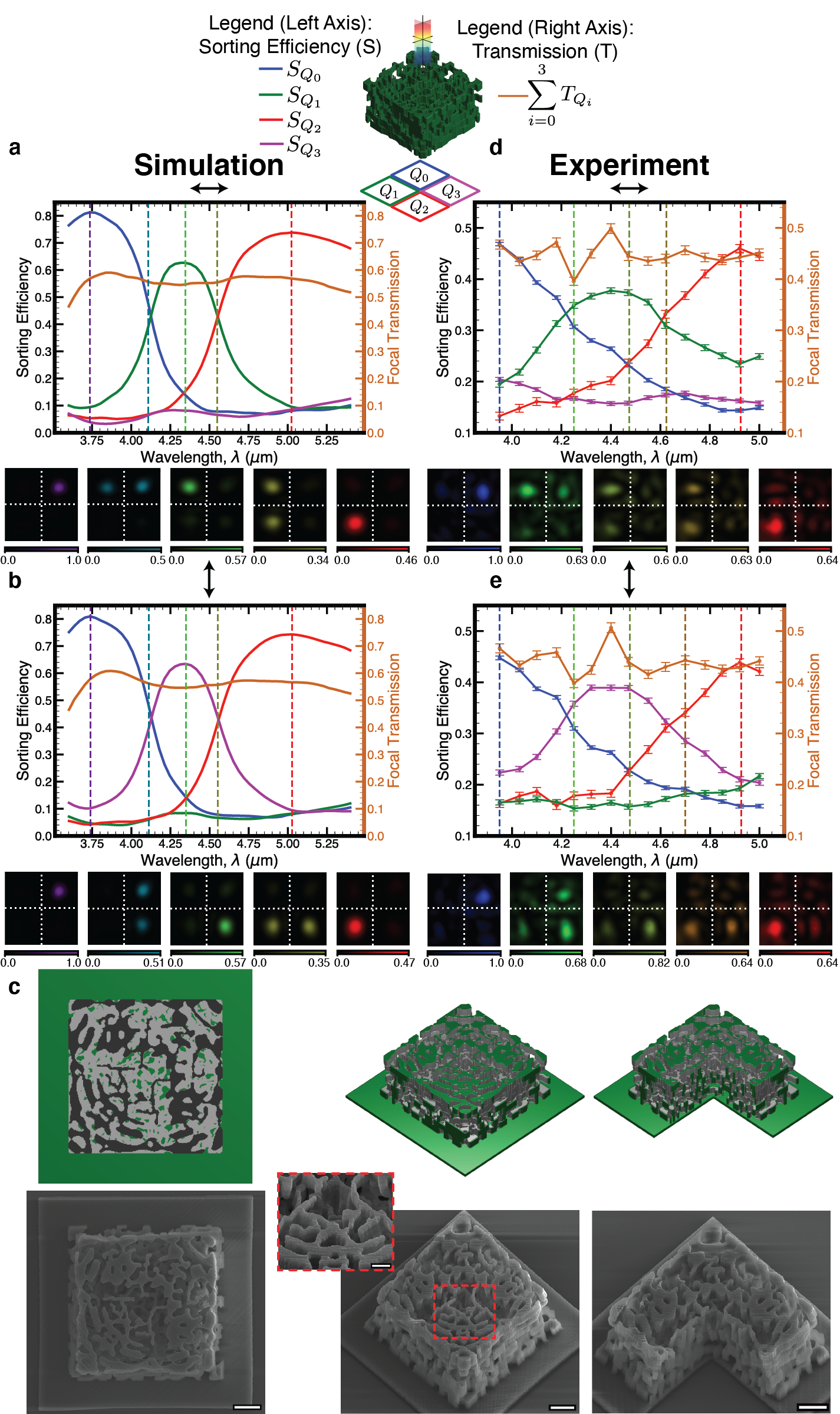"}
\end{center}
\caption{Fabrication and measurement results of multispectral and linear polarization sorting device.  \textbf{(a)} (top) Simulated device sorting spectrum showing both relative sorting efficiency ($S$) and net transmission ($T$) to focal plane normalized to pinhole transmission. For quadrant $k$, $S_{Q_k}=\frac{T_{Q_k}}{\sum_{i=0}^3{T_{Q_i}}}$. (bottom) Colored intensity images account for the expected imaging lens numerical aperture (NA=0.67) and show the focal spot moving as a function of wavelength.  Each color corresponds to the same colored dashed vertical line in the spectrum above it.  Intensity units are arbitrary, but comparable between all plots in (a).  Different maximum values on the colorbars here and in other figures are labeled and utilized for better visibility of the plotted intensity features.  \textbf{(b)} Same plots as in (a) for vertical polarization input.  \textbf{(c)} Schematic and associated SEM images of fabricated devices.  The rightmost device was printed with one quarter missing to show internal structure.  Scale bars: $5\mu m$, $5 \mu m$ (inset $2 \mu m$), $5 \mu m$ from left to right.  \textbf{(d, e)} Experimental comparison plots to (a, b) respectively with standard deviation (SD) error bars.  Note the differences between x- and y-axes offset and scale when comparing to simulation.  Wavelength range differences for the x-axis are due to the limited tuning range of the QCL used experimentally.}
\end{figure}\label{fig2}

\begin{figure}
\thisfloatpagestyle{plain}
\begin{center}
    \includegraphics{"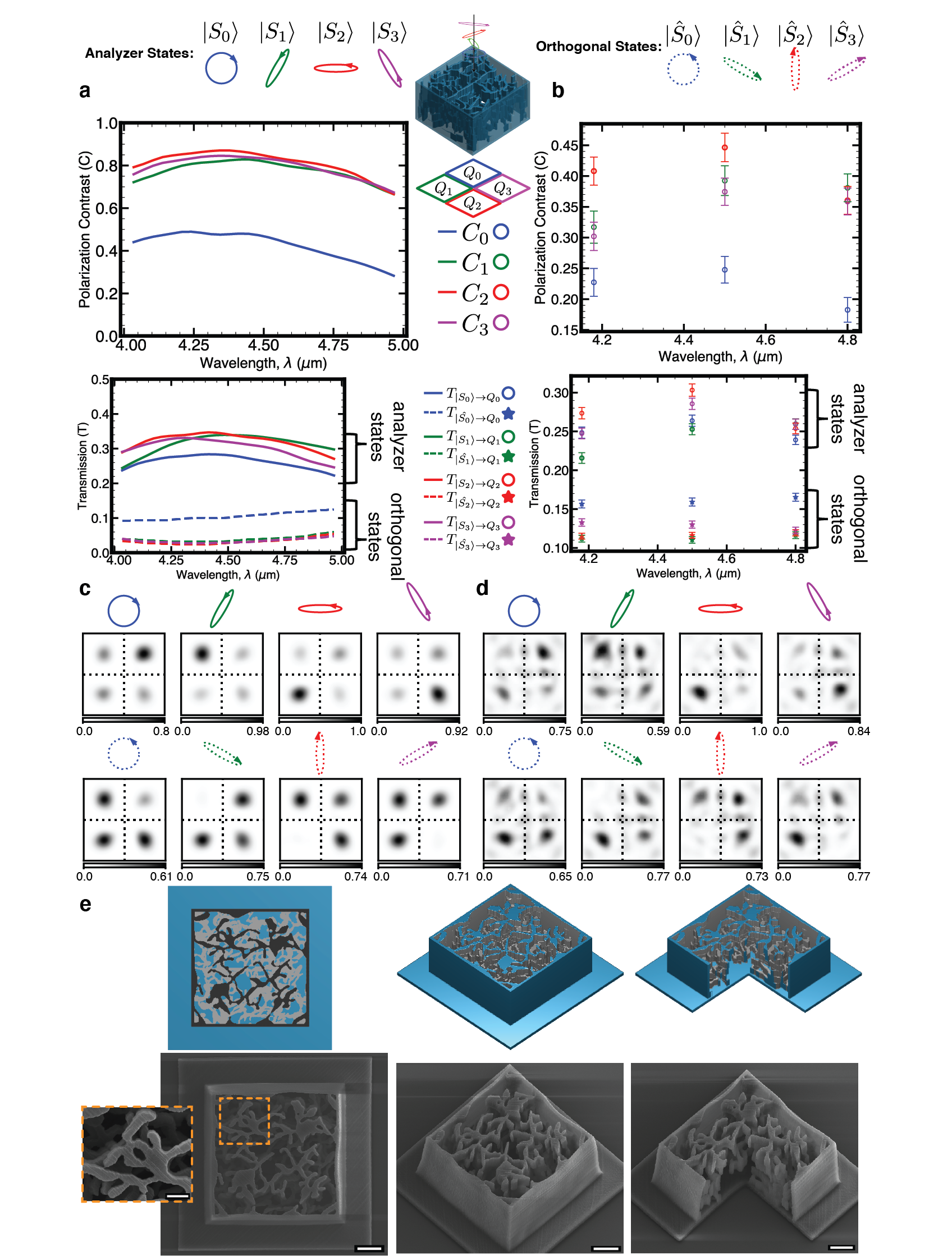"}
\end{center}
\caption{Fabrication and measurement results of Stokes polarimetry device.  \textbf{(a)} Polarization contrast ($C$) in simulation quantifying the transmission ($T$) into the desired quadrant for a given analyzer state versus transmission into the same quadrant for the orthogonal state.  For input $k$, $C_k=\frac{T_{\ket{S_k} \rightarrow Q_k } - T_{\ket{\hat{S_k}} \rightarrow Q_k}}{T_{\ket{S_k} \rightarrow Q_k } + T_{\ket{\hat{S_k}} \rightarrow Q_k}}$. Below is the transmission into the desired quadrants for the analyzer states (solid) and their orthogonal complements (dashed).  \textbf{(b)} Comparison plot of contrast and transmission for the experimental results with analyzer states shown with open circles and orthogonal states shown with stars in the transmission plot (SD error bars).  \textbf{(c)} Simulated focal intensity images ($\lambda=4.5\mu m$) accounting for imaging lens numerical aperture (NA=0.67) for the various input states where the top row contains analyzer states and the bottom row contains orthogonal states. Intensity units are arbitrary, but comparable between all plots in (c). \textbf{(d)} Experimental focal intensity images ($\lambda=4.5\mu m$) showing a bright quadrant for each analyzer state and the same quadrant dark for the complementary orthogonal state.  Intensity units are arbitrary, but comparable between all plots in (d).  \textbf{(e)} Schematic and associated SEM images of fabricated devices.  The rightmost device was printed with one quarter missing to show internal structure.  Scale bars: $5\mu m$ (inset $2 \mu m$), $5 \mu m$, $5 \mu m$ from left to right.}
\end{figure}\label{fig3}


\begin{figure}
\thisfloatpagestyle{plain}
\begin{center}
    \includegraphics{"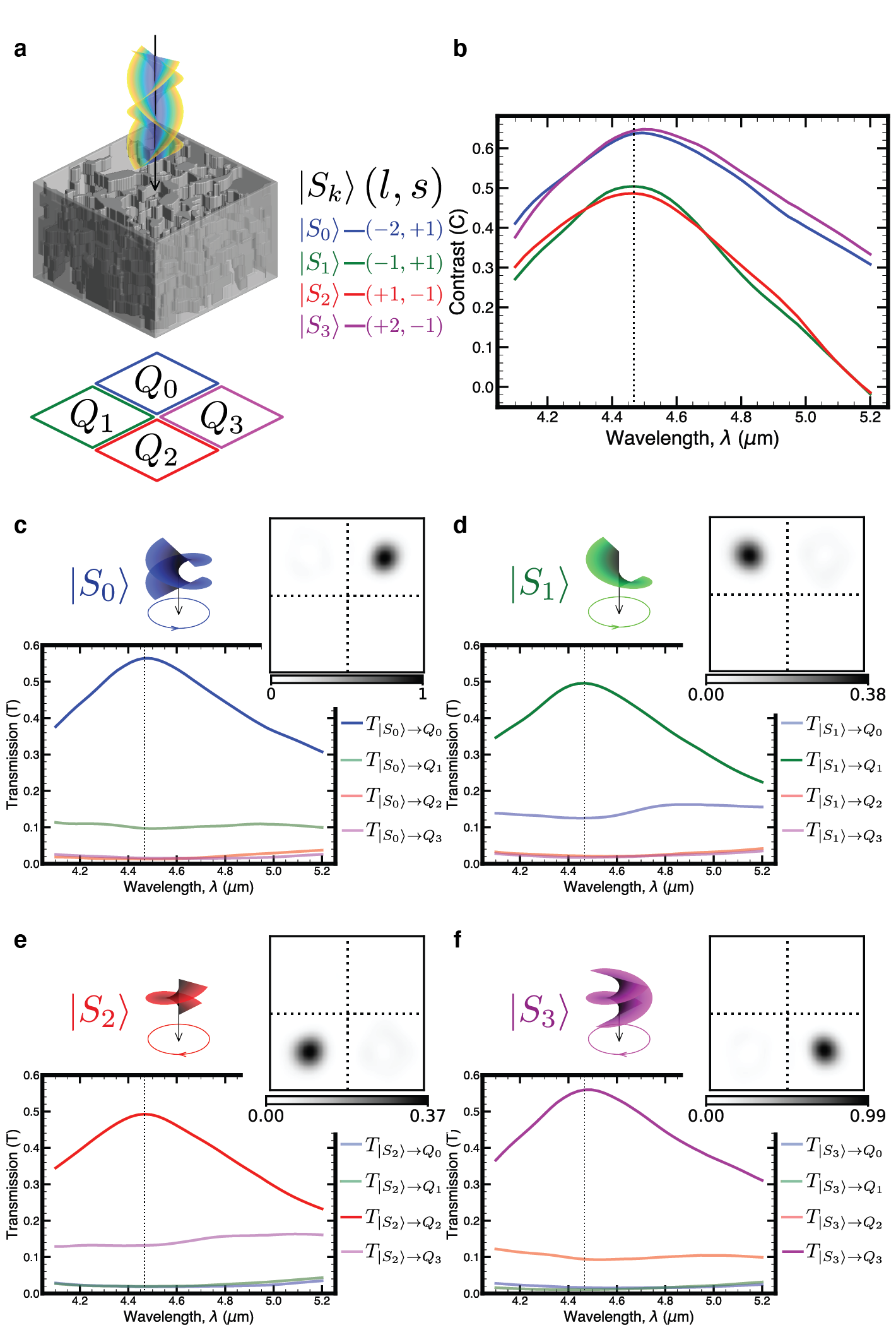"}
\end{center}
\caption{Simulation performance for angular momentum sorting device. \textbf{(a)} Schematic of device and focal plane quadrants. \textbf{(b)} Contrast for sorting each state ($C\in[-1, 1]$) defined by the transmission of a state into the desired quadrant versus the transmission into the rest of the focal plane.  For source $k$, $C_k = \frac{T_{\ket{S_k} \rightarrow Q_k } - \sum_{i \neq k }{ T_{\ket{S_k} \rightarrow Q_i}}}{T_{\ket{S_k} \rightarrow Q_k } + \sum_{i \neq k }{ T_{\ket{S_k} \rightarrow Q_i}}}$. \textbf{(c)} Transmission spectrum for ($l=-2$, $s=+1$) input with desired quadrant transmission in blue.  Transmission is normalized by power through the device aperture with no device present.  Inset: Intensity at focal plane (arbitrary units, but comparable to other intensity plots in figure). \textbf{(d-f)} Same plots as (c), but for ($l=-1$, $s=+1$), ($l=+1$, $s=-1$), ($l=+2$, $s=-1$), respectively.}
\end{figure}\label{fig4}

\newpage


\section*{Methods}

\subsection*{Simulation}
Simulations and inverse design optimizations of structures were carried out using Lumerical (ANSYS, Inc.) finite-difference time-domain (FDTD) Maxwell equations solver.  Working in the time domain with pulsed sources, the broadband response of the device to forward and adjoint sources can be computed in single simulations.  During design, we used a total field scattered field (TFSF) source to create a finite-sized, normally incident plane wave input to the device, depicted in \verb+Fig. S4+.  The simulation boundary conditions were perfectly matched layers (PML) to create the effect of an infinite simulation domain for the isolated structure. 

When verifying and reporting simulation performance in the manuscript, we doubled the mesh resolution in the device region, increased the simulation mesh accuracy everywhere from $2$ to $4$ (with $1$ being the least accurate and $8$ considered the most accurate), and used a defocused Gaussian source that matched more closely to the experiment compared to the plane wave optimization source.  We used a Gaussian beam with a waist radius of $w_0=12.5\mu m$ and a defocus amount at the aperture opening of $z=500\mu m$ such that the beam was diverging.  For the angular momentum device which we did not validate experimentally, we used focused Laguerre-Gaussian modes as defined below with their waist positioned at the device face for both optimization and evaluation.  We simulated devices on top of $Al_2O_3$ substrates using material permittivity values from literature \cite{LumericalMaterialDatabase}.  Further, for the multispectral and Stokes polarimetry we included the $30 \mu m$ diameter aperture in the optimization and evaluation using a perfect electrical conductor (PEC).

\subsection*{Simulation Resources}
Device optimizations are run on a high performance computing cluster.  During each optimization iteration, multiple forward and adjoint simulations are run to compute gradient information for the multi-objective problems specified below.  For reference, we used 10 computing nodes for the multispectral and Stokes polarimetry devices and 12 computing nodes for the angular momentum device.  Each node was allocated 8 Intel CPU cores (mix of Skylake 2.1 GHz and Cascadelake 2.2 GHz processors), to run simulations in parallel.  With this, the optimizations complete in approximately 30-40 hours of compute time depending on the specific device thickness in the paper.  The thicker Stokes polarimetry device shown in the supplemental took roughly 78 hours to complete.

\subsection*{Optimization Figures of Merit and Weighting}

\subsubsection*{\textbf{Multispectral and Linear Polarization}}
\begin{equation}
\begin{aligned}
    \max_{\Vec{\epsilon} \in \{\epsilon_{min}, \epsilon_{max}\}^N} g(\Vec{E}) = \sum_{\lambda} S((\sum_p \sum_q \kappa(q,p,\lambda) \frac{I_p(\Vec{r_q}, \lambda)}{I_{max}(\lambda)}); k) \\
    I_p(\Vec{r_q}, \lambda) = ||\Vec{E_p}(\Vec{r_q}, \lambda)||^2
\end{aligned}
\end{equation}
where $S(x;k)=\frac{ln(1 + e^{kx})}{k}$ is the softplus function, which ensures positivity of all figures of merit.  For large $x$, which corresponds to focusing of a given wavelength primarily in the correct quadrant, this function acts in a linear regime with the onset of this regime controlled by the value $k$.  In the opposite extreme, when a given wavelength is primarily in the undesired quadrants, $x$ is negative and the figure of merit tapers to zero ensuring its gradient contribution is large in the dynamic performance weighting scheme.  We use $k=2$ for this optimization.

The polarization index, $p$, separates performance for plane wave excitations with different linear polarizations.  The weighting $\kappa(q,p,\lambda)$ directs bands of wavelengths evenly spaced between $3.7-5\mu m$ to different quadrants, $q$, with the middle band sent to one of two locations based on linear polarization and the other two bands operating independent of polarization.  Wavelengths outside of a given band for a target quadrant are punished through a negative weight using $\kappa(q,p,\lambda)$.

\begin{align}
\kappa(q,p,\lambda) =
\begin{cases}
    1,& \text{if } (\lambda, p) \text{ desired for quadrant } q\\
    -\alpha,& \text{if } \lambda \text{ within } \Delta \lambda = \beta \frac{\Lambda}{3} \text{ of quadrant } q \text{'s target band}\\
    0,& \text{otherwise}
\end{cases}
\end{align}
Here, $\Lambda$ is the full bandwidth of the optimization, in our case $1.3\mu m$.  For our optimization, we used $\beta=0.5$, which controls how far spectrally into the neighboring bands we explicitly reject intensity in a given quadrant.  Finally, $\alpha$ sets the punishment weighting term for out-of-band light versus desired in-band light.  We used $\alpha=0.75$.

The normalization term $I_{max}(\lambda)$ accounts for the fact that for a fixed power, the intensity at the center of a focal spot will scale with wavelength.  We use the scaling $I_{max}(\lambda;l,f)=l^2/(f^2 \lambda^2)$ for device lateral size, $l$, and focal length, $f$.

\subsubsection*{\textbf{Full Stokes Polarimetry}}
In the Stokes polarimetry device, we searched for devices with high contrast in intensity in each quadrant when illuminated with an analyzer state versus the state orthogonal to it.  For each analyzer state ($a$) and its orthogonal state ($\bar{a}$), we measured the intensity in the middle of the quadrant:
\begin{equation}
\begin{aligned}
g_{a,q} = \frac{I_{a,q}(\Vec{r}_q, \lambda)}{I_{max}(\lambda)}\\
g_{\bar{a},q} = \frac{I_{\bar{a},q}(\Vec{r}_q, \lambda)}{I_{max}(\lambda)}\\
I_{a,q} = ||\vec{E}_{a,q}(\Vec{r}_q, \lambda)||^2\\
I_{\bar{a},q} = ||\vec{E}_{\bar{a},q}(\Vec{r}_q, \lambda)||^2
\end{aligned}
\end{equation}
We would like the analyzer value to be large and the other orthogonal value to be small, so we combined them together with the following product figure of merit:
\begin{align}
    g_q(\lambda)=g_{a,q}(\lambda)*(1-g_{\bar{a},q}(\lambda))
\end{align}

Further, given the bound of 50$\%$ transmission for an analyzer state to a quadrant for a perfect device, we only optimized the parallel intensity up until that point. The net figure of merit, then, is:
\begin{align}
    \max_{\Vec{\epsilon} \in \{\epsilon_{min}, \epsilon_{max}\}^N} g(\Vec{E}) = \sum_{\lambda} \sum_q g_q(\lambda)
\end{align}

\subsubsection*{\textbf{Angular Momentum}}

\begin{equation}
\begin{aligned}
    \max_{\Vec{\epsilon} \in \{\epsilon_{min}, \epsilon_{max}\}^N} g(\Vec{E}) = \sum_q \sum_s \sum_{\lambda} \kappa(q,s) \frac{I_s(\Vec{r_q}, \lambda)}{I_{max}(\lambda)} \\
    I_s(\Vec{r_q}, \lambda) = ||\Vec{E_s}(\Vec{r_q}, \lambda)||^2
\end{aligned}
\end{equation}
The weighting function $\kappa(q,s)$ controls whether focusing into a given quadrant, $q$, is desirable for a given mode source $s$.  For a total number of optimization iterations, $M$, we define $\kappa(q,s;m)$ parameterized by iteration number, $m$, as

\begin{align}
\kappa(q,s;m) =
\begin{cases}
    1,& \text{if } q=s\\
    -w_{end},& \text{if } m \geq m_{end}, q\neq s\\
    -(w_{start} + (w_{end} - w_{start})\frac{m}{m_{end}-1}),& \text{if }m < m_{end}, q\neq s
\end{cases}
\end{align}
We chose $w_{start}=\frac{1}{3}$, $w_{end}=1$, $m_{end}=90$, and $M=300$.  The first case corresponds to one quadrant being excited for a given source, so it receives a positive weight.  The second two cases describe a linear ramp of negative weight for rejecting intensity into incorrect quadrants for a certain number of iterations after which point a constant negative weight is used for the rest of the optimization.  With the dynamic performance weighting function described below, it is important that individual figures of merit stay positive for the entire optimization.  The ramping of the negative weighting helps in this regard and our specific optimization maintains positive individual figures of merit throughout.  We emphasize this is not a guarantee of the weighting scheme above, but a fortunate instance where it worked for our optimization.  In other optimizations, we explicitly ensured the individual figures of merit remained positive.

\subsubsection*{\textbf{Dynamic Weighting}}
All optimizations are multi-objective and require balancing many individual figures of merit.  One way of achieving a balance and preventing certain figures of merit from dominating the optimization solution is by using a dynamic performance-based weighting scheme.  As certain figures of merit start performing better than others, their relative optimization weight decreases.  For $N_{FOM}$ individual figures of merit, the $j^{th}$ figure of merit with current performance $f_j$ (with net performance defined as $\sum_j f_j$) is weighted:
\begin{align}
    w_j = \frac{2}{N_{FOM}} - \frac{f_j ^p}{\sum_n f_n ^p}
\end{align}

For the multispectral and linear polarization optimization, $f_j$ corresponded to each wavelength figure of merit.  In other words, $f_j = f_{\lambda} = S((\sum_p \sum_q \kappa(q,p,\lambda) \frac{I_p(\Vec{r_q}, \lambda)}{I_{max}(\lambda)}); k)$.
In the Stokes polarimetry case, $f_j$ corresponded to each product figure of merit for combinations of quadrant and wavelength.  Specifically, each $g_q(\lambda)$ was an individual figure of merit in the weighting scheme.
Finally, for the angular momentum optimization, each quadrant figure of merit corresponded to an $f_j$.  In other words, $f_j = f_q = \sum_s \sum_{\lambda} \kappa(q,s) \frac{I_s(\Vec{r_q}, \lambda)}{I_{max}(\lambda)}$.

We chose $p=2$ for all optimizations.  This weighting scheme relies on positive individual figures of merit to function properly.  With the above formula, weights can become negative when a given figure of merit is far ahead of the others.  In these cases, we simply shifted and rescaled to ensure all weights were greater than or equal to $0$.  After computing individual gradients for each figure of merit, $\frac{\partial f_j}{\partial \Vec{\epsilon}}$, the net gradient was computed using this weighting scheme as $\sum_j w_j \frac{\partial f_j}{\partial \Vec{\epsilon}}$.

The gradients derived during the optimization were based on a focusing figure of merit (norm electric field squared at a point) and use dipole excitations as adjoint sources.  Typically, this is a good proxy for transmission into a given quadrant, which is ultimately what we care to achieve with devices placed on top of focal plane arrays.  For some optimizations, the transmission-measured performance is used instead of the intensity-measured performance as the input to the dynamic weighting function.  This reduced dependencies on exact normalizations of intensity with power and took into account electric field profiles that looked different than a traditional focusing profile or were shifted from the center of the quadrant.

\subsection*{Optimization Fabrication Constraints}
Projection filters were used to push the optimization solution towards devices that respected certain fabrication constraints \cite{ProjectionMethods}.  These filters are differentiable functions applied in sequence to a design variable in order to create a device variable.  The device variable describes the structure being optimized and that will eventually be fabricated.  We found the gradient of the figure of merit with respect to the device variable and then backpropagated that gradient to the design variable using the chain rule.  The design variable was then stepped in the gradient direction.

For binarization, we used a sigmoid filter of the form $f(\rho_k) = \frac{tanh(\beta \eta) + tanh(\beta(\rho_k - \eta))}{tanh(\beta \eta) + tanh(\beta (1 - \eta))}$.  The strength, $\beta$, was increased over a series of $10$ epochs, each with $30$ iterations, starting at $\beta=0.0625$ and doubling each epoch.  The center point was fixed at $\eta=0.5$.  For layering, permittivity values were averaged vertically over the layer thickness at each lateral point.  This corresponds to averaging the two-dimensional gradient of each slice in a given layer during backpropagation.  

For minimum feature size, we used an averaging blurring function that tapered from the center pixel, specifically:
\begin{align}
    f(\rho_k) = \frac{1}{N}\sum_{r_j \in \Omega_k}\frac{b_r + 1 - r_j}{b_r}\rho_j
\end{align}
where $N=\sum_{r_j \in \Omega_k}\frac{b_r + 1 - r_j}{b_r}$ for normalization \cite{FeatureBlurringFilter}.  $b_r$ is the blur radius based on a circle inscribed on a square with sides equal to the desired minimum feature size of $750nm$, such that $b_r = 0.5 * \sqrt{2} * 750nm = 530nm$ ($r_j$ is the distance of voxel $j$ from voxel $k$).  We computed this sum out to $r_{j,max} = b_r$.  This filter tends to increase the density value of voxels nearby a solid one.  This encourages a minimum feature size in the solid domain.  However, the drawback of this method is that it does not guarantee a minimum feature size and we do end up with features smaller than the minimum value in our final designs.  This can be improved through filters that blur even further out, directly fixing the design grid to be that of minimum feature size increments, or use of a level set procedure at the end of the optimization with a feature size constraint.
The other disadvantage of this method is it only attempts to control feature size in the solid domain and does not address minimum gap sizes in the void domain.  

Some fabrication constraints are difficult to encapsulate in a filtering function.  Final designs needed to consist of a single piece of material for fabrication.  Further, with the Nanoscribe, an enclosed void cannot be realized because the liquid polymer would have no way to escape during development.  During the optimization, every $8$ iterations, each design layer was patched to ensure it was a single piece of material and bridges formed in this step were restricted from changing for the next $8$ iterations until the patching occurred again.  The bridges were chosen via a shortest path (Dijkstra's) graph algorithm with the cost equal to the amount a given voxel would need to move to become fully solid.  Islands of material were connected via a greedy minimum spanning tree approach with net bridge costs used as the edge weight for connecting two islands together.  The density value of the inserted bridge was then set to $0.75$ (fully solid corresponds to a density of $1$) everywhere along its path.  This works to ensure solid connectivity and often the void connectivity is maintained by chance throughout the optimization.  Final patches were done after the optimization to create void connectivity if it had not happened naturally.  These were usually minimal changes that did not have large effects on the device performance.  Nevertheless, the reported simulation results in this manuscript used the fully patched designs that were fabricated.

\subsection*{Device Fabrication}
The fabrication procedure for the measured devices is shown in \verb+Fig. S5+.  Devices were printed directly on top of 30$\mu m$ apertures defined via a photolithography-based lifotff procedure.  Apertures were 150$nm$ thick and controlled the illumination on the device.  They were also used for measuring beam power through blank apertures to normalize net transmission of the device. Negative-tone photoresist, AZ nLoF 2070 (MicroChemicals GmbH), was patterned with photolithography to create a variety of apertures as well as alignment marks for the optical setup.  Following oxygen and argon direct plasma cleaning to remove undesired residual photoresist left after development, 150$nm$ of aluminum (Al) was deposited via electron beam evaporation.  The apertures were lifted off in acetone and the substrate was cleaned in IPA followed by DI water.  IP-Dip resist was dropped onto the substrate surface for direct write lithography using the Nanoscribe Photonic Professional GT.  In the Nanoscribe, the apertures were located by moving the stage after the substrate surface was found by the microscope.  By turning on the laser below the polymerization threshold such that the microscope could still image a fluorescence signal from the laser focus, we aligned the center of the printing axes to the aperture center.  After writing, the devices were developed in propylene glycol methyl ether acetate (PGMEA) for 20 minutes and rinsed in two successive IPA baths for 3 minutes each.  The surface of the substrate was dried with a gentle nitrogen stream.  We found that critical point drying was not necessary for the integrity of our structures through the drying process.  For imaging in the scanning electron microscope, a 5$nm$ coating of platinum (Pt) was sputtered onto the surface to reduce charging effects due to the insulating polymer.

\subsection*{Optical Experimental Setup}
The optical setup shown in \verb+Fig. S1+ illuminated the devices through the sapphire substrate with a diverging Gaussian beam across the pinhole aperture on which the the device was printed.  Polarization in the case of the multispectral device was changed via a half-wave plate and linear polarizer where the former served to rotate more power into the polarization less overlapped with the laser output mode.  In the case of the Stokes measurement setup, a combination of a linear polarizer, half-wave plate and quarter-wave plate were used to achieve desired input states to the device.  These wave plates applied a wavelength-dependent retardance, which was taken into account in their chosen rotation.  To ensure the polarization states were correct, we used a traditional method for reconstruction of the polarization state consisting of a quarter-wave plate followed by a Wollaston prism (polarizing beam splitter).  The prism split orthogonal linear polarizations into two different angles which were imaged onto a power meter.  By rotating the quarter-wave plate to three known rotations and measuring the power in each angle, we reconstructed what the input polarization state must have been.  The reconstructed state overlaps are shown in \verb+Fig. S2c+ for each probing wavelength of the Stokes measurement setup.

\subsubsection*{\textbf{Imaging, Focal Length, and Chromatic Dispersion}}
The imaging objective, $L2$ in \verb+Fig. S1+, was translated in the axial direction to measure different planes moving back from the substrate surface.  First, we observed and verified in simulation the presence of significant focal shift with respect to the focal length of the device with wavelength due to chromatic aberration of the aspheric lens.  Modeling the optical setup in simulation using the Stellar Software Beam4 open source ray tracing program, we found this shift to be $31.4 \mu m$ between $\lambda = 3.95\mu m$ and $\lambda = 5\mu m$.  Experimentally, we measured this effect by imaging the diffraction pattern of an empty $30 \mu m$ aperture on the substrate surface.  By tuning the focus until the diffraction pattern disappeared, we could ensure we were focused on the aperture surface for a given wavelength.  We adjusted the axial position of the lens for different laser wavelengths until we were at the surface of the aperture and noted the micrometer position in order to characterize the chromatic focal shift of the imaging lens.  Experimentally, we computed this dispersion to be $46 \mu m$ for the same range of $\lambda = 3.95\mu m$ and $\lambda = 5\mu m$.  If we assume this can be used as a calibration of the micrometer on the stage, then each marking corresponds to $\frac{0.68\mu m}{tick}$.

Using stage markings, we found the best focal plane of the multispectral device for $\lambda = 3.95\mu m$ to be located at $63$ ticks from the substrate surface.  For a total device height of $19.5 \mu m$ and designed focal length of $25 \mu m$, we expected the focus to be located at $44.5 \mu m$ off the substrate surface.  Applying the calibration above of $\frac{0.68\mu m}{tick}$, we estimate the measured focal plane to be located at $43\mu m$ from the surface ($f=23.5 \mu m$ for assumed device height of $19.5 \mu m$), which is close to the design and within reasonable inaccuracies of the above calibration and small errors in printed device height.

For the multispectral device, we took $15$ measurements evenly spaced between $3.95\mu m$ and $5 \mu m$.  Since the chromatic dispersion is not equal across this whole range, we broke the range into two parts and linearly interpolated the axial position of the imaging objective to probe the same focal plane for each wavelength.  Between $3.95\mu m$ and $4.48 \mu m$, we interpolated over $26$ ticks corresponding to $17.7 \mu m$ and between $4.48 \mu m$ and $5 \mu m$, we interpolated over $20$ ticks corresponding to $13.7 \mu m$.  

For the Stokes polarimetry device, we measured at three distinct wavelengths, $4.18 \mu m$, $4.5 \mu m$, and $4.8 \mu m$.  We directly set the focal lengths based on the empirical axial position of a blank $30\mu m$ aperture on the substrate for each wavelength.  We used the same focal length of $63$ ticks corresponding to around $43 \mu m$.  The device height in this case was designed to be $19.8 \mu m$, so this corresponded to $f = 23.2 \mu m$.  

\subsubsection*{\textbf{Transmission Normalization}}
The imaging of device focal planes onto the camera needed to be calibrated with a net device transmission.  We quantified the transmission of the device by using an empty circular aperture on the substrate with the same diameter as the one sectioning off illumination to the device.  Using a power meter, we measured the power through the aperture without the device versus the power through the aperture with the device.  We tracked the beam center on the camera and the method depended on properly centering the beam on both apertures.  Further, we assumed that laser power was not fluctuating significantly in time and the beam center was not shifting upon successive wavelength tuning due to thermal effects and a changing laser mode.  Using this method, we saw consistent and expected transmission values through the device with only minor fluctuations for the multispectral device around $4.4 \mu m$, which may have been due to invalidation of assumptions previously stated.  This wavelength is close to the crossover between two modules in the QCL covering different spectral ranges and we speculate the power may be less stable here compared to other wavelengths.  The measured transmission was assumed to be contained in the camera image of the focal plane and its surrounding area.  We then assumed that the transmission corresponding to a patch of the camera image was equal to the ratio of its intensity to the total intensity multiplied by the net measured transmission.  Transmission into the focal plane, for example, was computed by multiplying the measured total transmission value by the ratio of intensity in the focal plane to the intensity in the focal plane and surrounding area.  Camera images were taken of the focal plane for each wavelength and a background of the camera (with the laser emission off) was taken immediately afterward.  Taking the background immediately after each measurement reduced error in the background drifting over the course of the long experimental procedure from temperature drifts in the room or the camera housing itself.  These background images were subtracted from the camera images.

\subsection*{Stokes State Creation and Verification}
Each polarization state was generated through choice of rotation of a linear polarizer, a half-wave plate, and a quarter-wave plate pictured in \verb+Fig. S2a+.  The wave plates are chromatic components with a retardance defined for a given wavelength.  Using the provided retardance data from Thorlabs for each component, we computed the effect a rotated component will have on each input wavelength.  By optimizing the choice of angles of these three components, we can generate all of the desired input polarization states to the device.  To verify the correctness of each state, we used the setup in \verb+Fig. S2b+ consisting of a quarter-wave plate and Wollaston prism.  The Wollaston prism splits the input polarization into its x- and y-polarization components, each of which are imaged onto and measured by a power meter.  Under different rotations of the quarter-wave plate, the magnitude of x- and y-polarized components will change as a function of the input state.  By measuring these components under three rotations and using the specified retardance values of this quarter-wave plate as a function of wavelength, we reconstructed the input state.  We used rotation values of $0 ^\circ$, $22^\circ$, and $44 ^\circ$ rotations of the fast axis with respect to the x-polarization direction. Plotted in \verb+Fig. S2c+ are the magnitudes of the vector overlaps of the reconstructed and the desired Jones state for each wavelength.  Note the y-axis on the plot begins at 0.9.

\section*{Acknowledgments}

We thank the staff at Northrop Grumman for allowing us to carry out mid-infrared measurements at their facility with their equipment and to Orrin Kigner for assistance with coordination as well as measurement advice.  Further, we acknowledge Chase Ellis from the U.S. Naval Resarch Laboratory and Professor Joe Tischler from University of Oklahoma for initial measurement and fabrication assistance as well as continued advice on the experimental setup and fabrication parameters. We thank Malina Strugaru for work on characterizing and verifying aspects of Nanoscribe fabrication, and Dr. Amir Arbabi, Dr. Oscar Bruno, Dr. Hyounghan Kwon and Ian Foo for helpful conversations and advice.

\section*{Declarations}

\subsection*{Funding}
Defense Advanced Research Projects Agency (HR00111720035)\\
Rothenberg Innovation Initiative (RI2), Caltech\\
Army Research Office (W911NF2210097)\\
Clinard Innovation Fund

\subsection*{Conflict of interest/Competing interests}
The authors have filed the following patents related to this work:
\t\\
Color and multi-spectral image sensor based on 3d engineered material (US20200124866A1)\\Broadband Polarization Splitting Based on Volumetric Meta-Optics (US20220004016)\\CMOS color image sensors with metamaterial color splitting (US11239276B2)

\subsection*{Availability of data and materials}
Data to support the conclusions in the manuscript can be provided on request.

\subsection*{Authors' contributions}
G.R., S.C.M., and A.F. conceived the project. G.R. carried out optimization, fabrication, and measurement of devices in the manuscript with input from other authors. C.B. consulted and provided critical feedback on optimization and measurement techniques and data analysis. T.Z. consulted on fabrication and optimization methods. J.C.G. and P.W.C.H. provided help with design and construction of experimental setup as well as invaluable consultation on measurement results, as well as hosted G.R. for measurements on many occasions at Northrop Grumman.  G.R. prepared the manuscript with input from all authors.

\bibliographystyle{unsrt}

\begin{thebibliography}{10}

\bibitem{RecentMetaReview}
Hou-Tong Chen, Antoinette~J Taylor, and Nanfang Yu.
\newblock {A review of metasurfaces: physics and applications}.
\newblock {\em Reports on progress in physics}, 79(7):076401, 2016.

\bibitem{NTTRecentRouter}
Masashi Miyata, Naru Nemoto, Kota Shikama, Fumihide Kobayashi, and Toshikazu
  Hashimoto.
\newblock {Full-color-sorting metalenses for high-sensitivity image sensors}.
\newblock {\em Optica}, 8(12):1596--1604, 2021.

\bibitem{MillerThesis}
Owen~Dennis Miller.
\newblock {\em {Photonic design: From fundamental solar cell physics to
  computational inverse design}}.
\newblock University of California, Berkeley, 2012.

\bibitem{MoleskyInvDesReview}
Sean Molesky, Zin Lin, Alexander~Y Piggott, Weiliang Jin, Jelena
  Vuckovi{\'{c}}, and Alejandro~W Rodriguez.
\newblock {Inverse design in nanophotonics}.
\newblock {\em Nature Photonics}, 12(11):659--670, 2018.

\bibitem{VuckovicSpectralDemultiplex}
Logan Su, Alexander~Y Piggott, Neil~V Sapra, Jan Petykiewicz, and Jelena
  Vuckovic.
\newblock {Inverse design and demonstration of a compact on-chip narrowband
  three-channel wavelength demultiplexer}.
\newblock {\em Acs Photonics}, 5(2):301--305, 2018.

\bibitem{YablanovitchInvDesFoundation}
Christopher~M Lalau-Keraly, Samarth Bhargava, Owen~D Miller, and Eli
  Yablonovitch.
\newblock {Adjoint shape optimization applied to electromagnetic design}.
\newblock {\em Optics express}, 21(18):21693--21701, 2013.

\bibitem{SellFanGrating}
David Sell, Jianji Yang, Sage Doshay, Rui Yang, and Jonathan~A Fan.
\newblock {Large-angle, multifunctional metagratings based on freeform
  multimode geometries}.
\newblock {\em Nano letters}, 17(6):3752--3757, 2017.

\bibitem{Arbabi2LayerMeta}
Mahdad Mansouree, Hyounghan Kwon, Ehsan Arbabi, Andrew McClung, Andrei Faraon,
  and Amir Arbabi.
\newblock {Multifunctional 2.5 D metastructures enabled by adjoint
  optimization}.
\newblock {\em Optica}, 7(1):77--84, 2020.

\bibitem{FaraonOpticalVolumetric}
Philip Camayd-Mu{\~{n}}oz, Conner Ballew, Gregory Roberts, and Andrei Faraon.
\newblock {Multifunctional volumetric meta-optics for color and polarization
  image sensors}.
\newblock {\em Optica}, 7(4):280--283, 2020.

\bibitem{NanoscribeStackedRefractive}
Timo Gissibl, Simon Thiele, Alois Herkommer, and Harald Giessen.
\newblock {Two-photon direct laser writing of ultracompact multi-lens
  objectives}.
\newblock {\em Nature photonics}, 10(8):554--560, 2016.

\bibitem{NanoscribeStackedDiffractive}
Simon Thiele, Christof Pruss, Alois~M Herkommer, and Harald Giessen.
\newblock {3D printed stacked diffractive microlenses}.
\newblock {\em Optics Express}, 27(24):35621--35630, 2019.

\bibitem{BraunScribe}
Christian~R Ocier, Corey~A Richards, Daniel~A Bacon-Brown, Qing Ding, Raman
  Kumar, Tanner~J Garcia, Jorik Van De~Groep, Jung-Hwan Song, Austin~J
  Cyphersmith, and Andrew Rhode.
\newblock {Direct laser writing of volumetric gradient index lenses and
  waveguides}.
\newblock {\em Light: Science {\&} Applications}, 9(1):1--14, 2020.

\bibitem{2DConcentrator}
Charles Roques-Carmes, Zin Lin, Rasmus~E Christiansen, Yannick Salamin,
  Steven~E Kooi, John~D Joannopoulos, Steven~G Johnson, and Marin
  Solja{\v{c}}i{\'{c}}.
\newblock {Toward 3D-Printed Inverse-Designed Metaoptics}.
\newblock {\em ACS Photonics}, 2022.

\bibitem{BayerPatent}
Bryce~E Bayer.
\newblock {Color imaging array}.
\newblock {\em United States Patent 3,971,065}, 1976.

\bibitem{PanasonicRouter}
Seiji Nishiwaki, Tatsuya Nakamura, Masao Hiramoto, Toshiya Fujii, and Masa-aki
  Suzuki.
\newblock {Efficient colour splitters for high-pixel-density image sensors}.
\newblock {\em Nature Photonics}, 7(3):240--246, 2013.

\bibitem{FanFreeformSplitters}
Nathan Zhao, Peter~B Catrysse, and Shanhui Fan.
\newblock {Perfect RGB‐IR Color Routers for Sub‐Wavelength Size CMOS Image
  Sensor Pixels}.
\newblock {\em Advanced Photonics Research}, 2(3):2000048, 2021.

\bibitem{ColorSplitterSims}
Eric Johlin.
\newblock {Nanophotonic color splitters for high-efficiency imaging}.
\newblock {\em Iscience}, 24(4):102268, 2021.

\bibitem{MIRFingerprintRef}
Albert Schliesser, Nathalie Picqu{\'{e}}, and Theodor~W H{\"{a}}nsch.
\newblock {Mid-infrared frequency combs}.
\newblock {\em Nature photonics}, 6(7):440--449, 2012.

\bibitem{DetectCO2}
Jane Hodgkinson, Richard Smith, Wah~On Ho, John~R Saffell, and Ralph~P Tatam.
\newblock {Non-dispersive infra-red (NDIR) measurement of carbon dioxide at 4.2
  {$\mu$}m in a compact and optically efficient sensor}.
\newblock {\em Sensors and Actuators B: Chemical}, 186:580--588, 2013.

\bibitem{MetamaterialAbsorbGas}
Sungho Kang, Zhenyun Qian, Vageeswar Rajaram, Sila~Deniz Calisgan, Andrea
  Al{\`{u}}, and Matteo Rinaldi.
\newblock {Ultra‐narrowband metamaterial absorbers for high spectral
  resolution infrared spectroscopy}.
\newblock {\em Advanced Optical Materials}, 7(2):1801236, 2019.

\bibitem{CancerDetectionMIR}
Angela~B Seddon.
\newblock {Mid‐infrared (IR)–A hot topic: The potential for using mid‐IR
  light for non‐invasive early detection of skin cancer in vivo}.
\newblock {\em physica status solidi (b)}, 250(5):1020--1027, 2013.

\bibitem{FTIRBioMaterials}
Matthew~J Baker, Júlio Trevisan, Paul Bassan, Rohit Bhargava, Holly~J Butler,
  Konrad~M Dorling, Peter~R Fielden, Simon~W Fogarty, Nigel~J Fullwood, and
  Kelly~A Heys.
\newblock {Using Fourier transform IR spectroscopy to analyze biological
  materials}.
\newblock {\em Nature protocols}, 9(8):1771--1791, 2014.

\bibitem{MIRPlasmonic}
Ang Wang and Yaping Dan.
\newblock {Mid-infrared plasmonic multispectral filters}.
\newblock {\em Scientific reports}, 8(1):1--7, 2018.

\bibitem{ThermalBlindness}
Huijie Zhao, Yansong Li, Guorui Jia, Na~Li, Zheng Ji, and Jianrong Gu.
\newblock {Comparing analysis of multispectral and polarimetric imaging for
  mid-infrared detection blindness condition}.
\newblock {\em Applied Optics}, 57(24):6840--6850, 2018.

\bibitem{SupplementaryReference}
{Materials, methods, and additional text are available in the supplementary.}

\bibitem{ProjectionMethods}
Fengwen Wang, Boyan~Stefanov Lazarov, and Ole Sigmund.
\newblock {On projection methods, convergence and robust formulations in
  topology optimization}.
\newblock {\em Structural and multidisciplinary optimization}, 43(6):767--784,
  2011.

\bibitem{IPDipIndex}
Daniel~B Fullager, Glenn~D Boreman, and Tino Hofmann.
\newblock {Infrared dielectric response of nanoscribe IP-dip and IP-L monomers
  after polymerization from 250 cm{\$}{\^{}}{\{}-1{\}}{\$} to 6000
  cm{\$}{\^{}}{\{}-1{\}}{\$}}.
\newblock {\em Optical Materials Express}, 7(3):888--894, 2017.

\bibitem{TPPAccuracy}
Xiaoqin Zhou, Yihong Hou, and Jieqiong Lin.
\newblock {A review on the processing accuracy of two-photon polymerization}.
\newblock {\em Aip Advances}, 5(3):030701, 2015.

\bibitem{CPCancerDiagnosis}
Britt Kunnen, Callum Macdonald, Alexander Doronin, Steven Jacques, Michael
  Eccles, and Igor Meglinski.
\newblock {Application of circularly polarized light for non‐invasive
  diagnosis of cancerous tissues and turbid tissue‐like scattering media}.
\newblock {\em Journal of biophotonics}, 8(4):317--323, 2015.

\bibitem{PolarimetryFacialRec}
Kristan~P Gurton, Alex~J Yuffa, and Gorden~W Videen.
\newblock {Enhanced facial recognition for thermal imagery using polarimetric
  imaging}.
\newblock {\em Optics letters}, 39(13):3857--3859, 2014.

\bibitem{DepthSensingPolCues}
Achuta Kadambi, Vage Taamazyan, Boxin Shi, and Ramesh Raskar.
\newblock {Polarized 3d: High-quality depth sensing with polarization cues}.
\newblock In {\em Proceedings of the IEEE International Conference on Computer
  Vision}, pages 3370--3378, 2015.

\bibitem{SkyMonitoringPol}
Nathan~J Pust and Joseph~A Shaw.
\newblock {Digital all-sky polarization imaging of partly cloudy skies}.
\newblock {\em Applied optics}, 47(34):H190--H198, 2008.

\bibitem{BioPolNavigation}
Daobin Wang, Huawei Liang, Hui Zhu, and Shuai Zhang.
\newblock {A bionic camera-based polarization navigation sensor}.
\newblock {\em Sensors}, 14(7):13006--13023, 2014.

\bibitem{RotatingStokes}
H~Gordon Berry, G~Gabrielse, and A~E Livingston.
\newblock {Measurement of the Stokes parameters of light}.
\newblock {\em Applied optics}, 16(12):3200--3205, 1977.

\bibitem{PolReviewPassive}
J~Scott Tyo, Dennis~L Goldstein, David~B Chenault, and Joseph~A Shaw.
\newblock {Review of passive imaging polarimetry for remote sensing
  applications}.
\newblock {\em Applied optics}, 45(22):5453--5469, 2006.

\bibitem{MicropolFiltering}
Jing Bai, Chu Wang, Xiahui Chen, Ali Basiri, Chao Wang, and Yu~Yao.
\newblock {Chip-integrated plasmonic flat optics for mid-infrared full-Stokes
  polarization detection}.
\newblock {\em Photonics Research}, 7(9):1051--1060, 2019.

\bibitem{EhsanPolarimetry}
Ehsan Arbabi, Seyedeh~Mahsa Kamali, Amir Arbabi, and Andrei Faraon.
\newblock {Full-Stokes imaging polarimetry using dielectric metasurfaces}.
\newblock {\em Acs Photonics}, 5(8):3132--3140, 2018.

\bibitem{PolCameraCapasso}
Noah~A Rubin, Gabriele D’Aversa, Paul Chevalier, Zhujun Shi, Wei~Ting Chen,
  and Federico Capasso.
\newblock {Matrix Fourier optics enables a compact full-Stokes polarization
  camera}.
\newblock {\em Science}, 365(6448):eaax1839, 2019.

\bibitem{LGModes}
Les Allen, Marco~W Beijersbergen, R~J~C Spreeuw, and J~P Woerdman.
\newblock {Orbital angular momentum of light and the transformation of
  Laguerre-Gaussian laser modes}.
\newblock {\em Physical review A}, 45(11):8185, 1992.

\bibitem{OAMCommReview}
Alan~E Willner, Kai Pang, Hao Song, Kaiheng Zou, and Huibin Zhou.
\newblock {Orbital angular momentum of light for communications}.
\newblock {\em Applied Physics Reviews}, 8(4):041312, 2021.

\bibitem{AMCommNatPhot}
Jian Wang, Jeng-Yuan Yang, Irfan~M Fazal, Nisar Ahmed, Yan Yan, Hao Huang,
  Yongxiong Ren, Yang Yue, Samuel Dolinar, and Moshe Tur.
\newblock {Terabit free-space data transmission employing orbital angular
  momentum multiplexing}.
\newblock {\em Nature photonics}, 6(7):488--496, 2012.

\bibitem{OAMMuxDemuxFibers}
Nenad Bozinovic, Yang Yue, Yongxiong Ren, Moshe Tur, Poul Kristensen, Hao
  Huang, Alan~E Willner, and Siddharth Ramachandran.
\newblock {Terabit-scale orbital angular momentum mode division multiplexing in
  fibers}.
\newblock {\em science}, 340(6140):1545--1548, 2013.

\bibitem{ScienceAMMultiplex}
Haoran Ren, Xiangping Li, Qiming Zhang, and Min Gu.
\newblock {On-chip noninterference angular momentum multiplexing of broadband
  light}.
\newblock {\em Science}, 352(6287):805--809, 2016.

\bibitem{ComputationalMetasurfaceInvDes}
Zin Lin, Charles Roques-Carmes, Raphaël Pestourie, Marin Solja{\v{c}}i{\'{c}},
  Arka Majumdar, and Steven~G Johnson.
\newblock {End-to-end nanophotonic inverse design for imaging and polarimetry}.
\newblock {\em Nanophotonics}, 10(3):1177--1187, 2021.

\bibitem{PolymerInversion}
Nicolas T{\'{e}}treault, Georg von Freymann, Markus Deubel, Martin
  Hermatschweiler, Fabian P{\'{e}}rez‐Willard, Sajeev John, Martin Wegener,
  and Geoffrey~A Ozin.
\newblock {New route to three‐dimensional photonic bandgap materials: silicon
  double inversion of polymer templates}.
\newblock {\em Advanced Materials}, 18(4):457--460, 2006.

\bibitem{ParallelTPP}
Vincent Hahn, Pascal Kiefer, Tobias Frenzel, Jingyuan Qu, Eva Blasco,
  Christopher Barner‐Kowollik, and Martin Wegener.
\newblock {Rapid assembly of small materials building blocks (voxels) into
  large functional 3D metamaterials}.
\newblock {\em Advanced Functional Materials}, 30(26):1907795, 2020.

\bibitem{CMOSPlasmonics}
Lingyu Hong, Hao Li, Haw Yang, and Kaushik Sengupta.
\newblock {Fully integrated fluorescence biosensors on-chip employing
  multi-functional nanoplasmonic optical structures in CMOS}.
\newblock {\em IEEE Journal of Solid-State Circuits}, 52(9):2388--2406, 2017.

\bibitem{SubtractivePhotonics}
Reza Fatemi, Craig Ives, Aroutin Khachaturian, and Ali Hajimiri.
\newblock {Subtractive photonics}.
\newblock {\em Optics Express}, 29(2):877--893, 2021.

\bibitem{StructuralColorReview}
Yuqian Zhao, Yong Zhao, Sheng Hu, Jiangtao Lv, Yu~Ying, Gediminas Gervinskas,
  and Guangyuan Si.
\newblock {Artificial structural color pixels: A review}.
\newblock {\em Materials}, 10(8):944, 2017.

\bibitem{ARChallenges}
Yun-Han Lee, Tao Zhan, and Shin-Tson Wu.
\newblock {Prospects and challenges in augmented reality displays.}
\newblock {\em Virtual Real. Intell. Hardw.}, 1(1):10--20, 2019.

\bibitem{LumericalMaterialDatabase}
Edward~D Palik.
\newblock {\em {Handbook of optical constants of solids}}, volume~3.
\newblock Academic press, 1998.

\bibitem{FeatureBlurringFilter}
James~K Guest.
\newblock {Topology optimization with multiple phase projection}.
\newblock {\em Computer Methods in Applied Mechanics and Engineering},
  199(1-4):123--135, 2009.

\end{thebibliography}

\section*{Supplementary Information}\label{sec12}

\subsection*{Two-Photon Polymerization (TPP) Accuracy}
Fabrication via TPP is a flexible and powerful method, but also has known challenges in printing accuracy \cite{TPPAccuracy}.  We observe shrinkage of the structure, which is dependent on the height of the layer from the substrate. Material printed on the bottom layer is not able to shrink from its printed size because it is physically adhered to the substrate.  The topmost layer is roughly $90\%$ of the desired lateral size and the bottom layer is close to the expected size. We also observe dilation of the smallest features in the design.  Designs were compensated for this effect by pre-eroding features in the STL file before printing.  Finally, the Nanoscribe had a mismatch between the feature size in each lateral direction.  This is not a limitation of TPP, but likely the result of astigmatism in the optical alignment of our specific tool.

\subsection*{Laguerre Gaussian Modes for Angular Momentum Splitter}
A spatially varying field can carry orbital angular momentum (OAM).  Discrete values of OAM, $l$, can be found in the Laguerre-Gaussian orthonormal basis for solutions of the paraxial wave equation \cite{LGModes}.  We used a simplified set with $p=0$, such that each mode was defined at its waist ($z=0$) with spatial profile in cylindrical coordinates:
\begin{align}
    u(r,\phi,z=0) = (\frac{r\sqrt{2}}{w_0})^{|l|} e^{\frac{-r^2}{w_0^2}} e^{-i l \phi}
\end{align}
where $w_0$ is the waist radius of the beam.  We chose $w_0=8.5\mu m$ to ensure the mode was confined to the device.  Transmission plots shown are geometrically normalized against the transmission of this beam through the device aperture with no device present.
We can further assign a spin angular momentum of the mode by choosing the handedness of its circular polarization.  The following pairs of OAM values $l$ and spin values $s$ were used in the optimization: $(l,s)=(-2, 1),(-1, 1), (1, -1), (2, -1)$.  These states were assigned to quadrants starting with the top right (blue) and moving counterclockwise (green, red, magenta).

\subsection*{$12$-Layer Stokes Polarimetry Device}
The polarimetry device in the main text consists of six $3 \mu m$ layers and struggles to achieve equal contrast for all four analyzer states with the circular polarization state lagging the others.  We speculate this may be due to lack of degrees of freedom in the thickness of device.  As a comparison, we optimize a thicker device consisting of twelve $3 \mu m$ layers to see if the solution will display better contrast for all analyzer states.  In \verb+Fig. S3+ we show the comparison of the thicker device to the original.  While the quadrant transmission per analyzer state is slightly reduced, the contrast metric is improved for the circular polarization state without sacrificing the other analyzer state contrasts.

\subsection*{Polarimetry Splitting Bounds}

We can model the Stokes polarimetry device as a linear system that projects an input Jones state describing the x- and y-polarized electric field components onto several analyzer states.  The Jones polarization is a 2-dimensional complex vector.  The four analyzer states for our device are specifically chosen Jones vectors.  In \verb+Fig. S6+, analyzer states correspond to $\ket{v_i}$, where $N=4$ for the device in the paper.  We assume the device outputs into four spatially distinct modes $\ket{w_k}$, such that we take them to be orthogonal ($\braket{w_i|w_k}=\delta_{ik}$).  Specifically, we model each output mode as a focused spot in a different quadrant of the focal plane and thus we assume the lack of spatial overlap implies orthogonality to a good approximation.  The functionality of the device is described by an operator $\hat{Q}$ where projection of an input state on each analyzer direction modulates the amplitude of an outgoing mode.  We write
\begin{align}
    \hat{Q} = \sum_i{\alpha_i \ket{w_i} \bra{v_i}}
\end{align}
\noindent{}Without loss of generality, we assume $\alpha_i$ is real.  Any complex phase can be included in output mode $\ket{w_i}$.

\subsubsection*{\textbf{Maximum transmission into each analyzer state}}

Next, we assume for simplicity that all states have the same projection efficiency, such that $\alpha_i=\alpha$.  The transmission bound will differ from the following if each state does not split at the same projection efficiency.  Consider an arbitrary state $\ket{a}$ and it's orthogonal complement $\ket{\Bar{a}}$.  The action of $\hat{Q}$ on $\ket{a}$ is
\begin{align}
    \hat{Q}\ket{a} = \alpha \sum_i{\ket{w_i} \braket{v_i|a}}
\end{align}
\noindent{}Taking the vector magnitude squared of the resulting state
\begin{align}
    \braket{a|\hat{Q}^{\dagger}\hat{Q}|a} = \alpha^2 \sum_{i,j}{\braket{w_i|w_j} \braket{v_j|a}\braket{a|v_i}}
\end{align}
\noindent{}Since $\braket{w_i|w_k}=\delta_{ik}$, the double sum reduces to
\begin{align}
    \braket{a|\hat{Q}^{\dagger}\hat{Q}|a} = \alpha^2 \sum_{i}{ \braket{v_i|a}\braket{a|v_i}} = \alpha^2 \sum_{i}{|\braket{a|v_i}|^2}
\end{align}
\noindent{}Following this pattern, we also have
\begin{align}
    \braket{\Bar{a}|\hat{Q}^{\dagger}\hat{Q}|\Bar{a}} = \alpha^2 \sum_{i}{ \braket{v_i|\Bar{a}}\braket{\Bar{a}|v_i}} = \alpha^2 \sum_{i}{ |\braket{\Bar{a}|v_i}|^2}
\end{align}
\noindent{}Due to energy conservation, we cannot have gained any magnitude through applying $\hat{Q}$ on the state so $\braket{a|\hat{Q}^{\dagger}\hat{Q}|a} \leq 1$ and $\braket{\Bar{a}|\hat{Q}^{\dagger}\hat{Q}|\Bar{a}} \leq 1$.  Summing these together, we get
\begin{align}
    \braket{a|\hat{Q}^{\dagger}\hat{Q}|a} + \braket{\Bar{a}|\hat{Q}^{\dagger}\hat{Q}|\Bar{a}} = \alpha^2 \sum_{i}{ ( |\braket{a|v_i}|^2 + |\braket{\Bar{a}|v_i}|^2 )} \leq 2
\end{align}
\noindent{}Because the Jones vector space is 2-dimensional, $\ket{a}$ and $\ket{\Bar{a}}$ form an orthonormal basis, so by definition $( |\braket{a|v_i}|^2 + |\braket{\Bar{a}|v_i}|^2 ) = 1$.  Thus, the sum simply becomes
\begin{align}
    \braket{a|\hat{Q}^{\dagger}\hat{Q}|a} + \braket{\Bar{a}|\hat{Q}^{\dagger}\hat{Q}|\Bar{a}} = N \alpha^2 \leq 2
\end{align}
\noindent{}If we assume $\alpha$ is the largest it can be, then $\alpha^2=\frac{2}{N}$.  For $N=4$ as is the case for the device in this manuscript, $\alpha^2=0.5$.  Thus, the maximum transmission we can achieve for each analyzer state into its output mode is $0.5$.

\subsubsection*{\textbf{Minimum overlap between analyzer states}}

Given a maximum transmission efficiency of $0.5$ for each analyzer state, we can set a minimum overlap for Jones vector analyzer states used in the splitter.  While the choice is not unique, a maximally spaced set of vectors will have a common mutual overlap.  Assume for our set of analyzer states,
\begin{align}
    |\braket{v_i|v_j}|^2 = 
    \begin{cases}
        1 \text{ if $i=j$} \\
        \beta^2 \text{ if $i \neq j$}
    \end{cases}
\end{align}
\noindent{}Sending in an analyzer state to the device
\begin{align}
    \hat{Q}\ket{v_k} = \alpha \sum_i{\ket{w_i} \braket{v_i|v_k}}
\end{align}
\noindent{}Taking the magnitude like before and using the orthogonality of the $\ket{w_i}$ states
\begin{align}
    \braket{v_k|\hat{Q}^{\dagger}\hat{Q}|v_k} = \alpha^2 \sum_{i}{ \braket{v_i|v_k}\braket{v_k|v_i}} = \alpha^2 \sum_{i}{|\braket{v_k|v_i}|^2}
\end{align}
\noindent{}Using the common overlap between states in the analyzer set and requiring that by energy conservation this magnitude squared is bound by $1$,
\begin{align}
    \braket{v_k|\hat{Q}^{\dagger}\hat{Q}|v_k} = \alpha^2 ( 1 + ( N - 1 )\beta^2) \leq 1
\end{align}
\noindent{}The relation between $\alpha$ and $\beta$, then is given by
\begin{align}
    \alpha^2 \leq \frac{1}{1 + ( N - 1 )\beta^2}
\end{align}
\noindent{}Suppose we specialize to the case where the transmission is maximized into each analyzer state ($\alpha^2=\frac{2}{N}$) and we have no lost transmission for any given analyzer state through the system ($\braket{v_k|\hat{Q}^{\dagger}\hat{Q}|v_k} = 1$).  Then,
\begin{equation}
\begin{aligned}
    \alpha^2 ( 1 + ( N - 1 )\beta^2) = 1 \\
    \frac{2}{N}( 1 + ( N - 1 )\beta^2) = 1 \\
    1 + ( N - 1 )\beta^2 = \frac{N}{2} \\
    (N - 1 )\beta^2 = \frac{ N - 2 }{ 2 } \\
    \beta^2 = \frac{N-2}{2(N-1)}
\end{aligned}
\end{equation}
\noindent{}Note the case of $N=2$ requires no overlap between the vectors with $\beta^2=0$ and $\alpha^2=\frac{2}{N}=1$ because that matches the dimensionality of the Jones vector space.  However, from two measurements, we cannot reconstruct the full Stokes vector where in order to do so we need at lease $N=4$.  As stated before, for $N=4$, $\alpha^2=0.5$ at best and with no lost transmission for the analyzer states, $\beta^2=\frac{1}{3}$.

\subsection*{Polarimetry Contrast Bounds}
The contrast figure of merit for the Stokes polarimetry device is independent of overall transmission.  For a given quadrant corresponding to analyzer state $\ket{v_i}$ and orthogonal complement $\ket{\Bar{v_i}}$, the contrast is related to the analyzer transmission $T_{analyzer}$ and orthogonal transmission $T_{orthogonal}$ to the quadrant as $C=\frac{T_{analyzer}-T_{orthogonal}}{T_{analyzer}+T_{orthogonal}}$.  In order to get a contrast of $C=1$, we need to be able to completely extinguish light in the analyzer quadrant for the orthogonal state.
\subsubsection*{\textbf{Analyzer state transmission to all quadrants}}
We first show that a given analyzer state must necessarily appear in more than just the desired quadrant.  Following from the notation above, the action of the device on an analyzer state, $\ket{v_k}$ is given by
\begin{align}
    \hat{Q}\ket{v_k} = \sum_i{\alpha_i \ket{w_i} \braket{v_i|v_k}}
\end{align}
\noindent{}We ask how much overlap does this have with one of the output modes $\ket{w_j}$ not corresponding to the analyzer quadrant (i.e. $i \neq j$).
\begin{align}
    \braket{w_j|\hat{Q}|v_k} = \sum_i{\alpha_i \braket{w_j|w_i} \braket{v_i|v_k}} = \alpha_j \braket{v_j|v_k}
\end{align}
\noindent{}where we used $\braket{w_j|w_i}=\delta_{ij}$ to eliminate the sum.  However, as we showed above, with four analyzer states, $\braket{v_j|v_k} \neq 0$ even for $j \neq i$.  So there is energy in the other quadrants according to the splitting efficiency of the $j^{th}$ analyzer state and the overlap between the $j$ and $k$ analyzer states.
\subsubsection*{\textbf{Extinguishing orthogonal state to analyzer quadrant}}
We now check if an orthogonal state can be completely extinguished to the analyzer quadrant, which will determine if we can achieve a contrast of $C=1$. When we send in the orthogonal state to a given analyzer, $\ket{\bar{v_k}}$, the device output is given by
\begin{align}
    \hat{Q}\ket{\bar{v_k}} = \sum_i{\alpha_i \ket{w_i} \braket{v_i|\bar{v_k}}}
\end{align}
\noindent{}Since it is true that $\braket{v_k|\bar{v_k}} = 0$ by definition, the sum is reduced to
\begin{align}
    \hat{Q}\ket{\bar{v_k}} = \sum_{i \neq k}{\alpha_i \ket{w_i} \braket{v_i|\bar{v_k}}}
\end{align}
\noindent{}Now, we ask how much overlap does this have with the output mode corresponding to this analyzer quadrant, $\ket{w_k}$, since we are interested in seeing if this overlap can be zero.
\begin{align}
    \braket{w_k|\hat{Q}|\bar{v_k}} = \sum_{i \neq k}{\alpha_i \braket{w_k|w_i} \braket{v_i|\bar{v_k}}} = 0
\end{align} where $\braket{w_k|w_i}=\delta_{ki}$ is only nonzero for $i = k$, but the sum explicitly ranges over values of $i \neq k$.  Thus, we can extinguish a quadrant completely for a given orthogonal state and a contrast of $1$ is theoretically achievable even if we transmit all incident light through the device to the focal plane.

\subsection*{Polarimetry Analyzer States}
The choice of analyzer states that fits the above criteria is not unique, but will correspond to a tetrahedron with points lying on the Poincaré sphere.  First, we choose evenly spaced pure polarization states in Stokes space and then evaluate their mutual overlaps in Jones space.  One state is fixed in Stokes space to be right circular polarzation (RCP), which is encoded as $\begin{bmatrix}1, & 0, & 0, & 1\end{bmatrix}$.  This choice is arbitrary and different starting states will generate equally suitable sets of analyzer states.  Staying on the Poincaré sphere surface means the first entry is fixed to $1$ (from here, we  write the vector in terms of $S_1, S_2,$ and $S_3$).  The other three states should lie on a circle with a fixed polar angle from this first state such that all mutual overlaps are the same.  For polar angle $\theta$ and azimuthal angle $\phi$, these states can be parameterized $\begin{bmatrix}\sin{\theta} \cos{\phi},& \sin{\theta} \sin{\phi}, & \cos{\theta}\end{bmatrix}$.  To evenly spread out these states azimuthally, the spacing should be $\Delta\phi=\frac{2\pi}{3}$.  We make the non-unique choice to set the first $\phi=0$.  The first two states on the circle, then are $\begin{bmatrix}\sin{\theta},& 0, & \cos{\theta}\end{bmatrix}$ and $\begin{bmatrix}\sin{\theta} \cos{\frac{2\pi}{3}},& \sin{\theta} \sin{\frac{2\pi}{3}}, & \cos{\theta}\end{bmatrix}$.  Evaluating the dot product between any of the states on the circle and the right circular polarization state yields $\cos{\theta}$.  The first two states on the circle have a dot product of $\sin^2{\theta}\cos{\frac{2\pi}{3}} + \cos^2{\theta}$.  Equating these two values generates the relation:
\begin{align}
    \sin^2{\theta}\cos{\frac{2\pi}{3}} + \cos^2{\theta} = \cos{\theta}
\end{align}
\noindent{}Solving for $\cos{\theta}$ gives $\cos{\theta} = -\frac{1}{3}$.  Completing the tetrahedron, the final Stokes states (rounded to the thousands place) are:
\begin{equation}
\begin{aligned}
\begin{bmatrix} 1, & 0, & 0, & 1 \end{bmatrix} \\
\begin{bmatrix} 1, & -0.471, & 0.816, & -0.333 \end{bmatrix} \\
\begin{bmatrix} 1, & 0.943, & 0, & -0.333 \end{bmatrix} \\
\begin{bmatrix} 1, & -0.471, & -0.816, & -0.333 \end{bmatrix}
\end{aligned}
\end{equation}
\noindent{}
Converting these states to Jones vectors, the analyzer states we used (rounded to the thousands place) are given by:
\begin{equation}
\begin{aligned}
\begin{bmatrix} 0.707, & -0.707j \end{bmatrix} \\
\begin{bmatrix} 0.514, & 0.794+0.324j \end{bmatrix} \\
\begin{bmatrix} 0.986, & 0.169j \end{bmatrix} \\
\begin{bmatrix} 0.514, -0.794 + 0.324j \end{bmatrix}
\end{aligned}
\end{equation}
\noindent{}The squared overlap magnitudes between any of these states, $\beta^2 = \frac{1}{3}$ as desired for equally split analyzer states.

\subsection*{Device Index of Refraction Profiles}
Optimized index of refraction profiles for the multispectral and angular momentum sorting devices are shown in \verb+Fig. S7+ and those for the Stokes polarimetry device from the main text and the one from the supplement with more layers are shown in \verb+Fig. S8+.

\subsection*{Polarimetry Reconstruction}

The following section shows how the polarimetry device presented in the main text can be used to recover the Stokes parameters of arbitrarily polarized inputs.  This addresses interpretation of quadrant outputs when the excitation is different than the four analyzer states used in the design.  It further addresses the ability of the device to utilize the four measurements to recover the degree of polarization for partially polarized light.  This exploration is done in simulation, but the same calibration and reconstruction procedure can be used experimentally as well.

\subsubsection*{\textbf{Reconstruction Method}}

The problem of converting the signal in each of the four quadrants into the incident polarization state can be phrased as follows:
\begin{equation}
    \underline{\underline{M}}\vec{S} = \vec{T}
\end{equation}
where $\underline{\underline{M}}$ is the forward model that maps the Stokes vector, $\vec{S}$, to the observed quadrant transmissions, $\vec{T}$.  We utilize the common definition of the Stokes parameters:
\begin{equation}
    \vec{S} = \begin{bmatrix}S_0\\S_1\\S_2\\S_3\end{bmatrix}=\begin{bmatrix}E_x^2 + E_y^2 = E_{45}^2 + E_{-45}^2 = E_R^2 + E_L^ 2\\E_x^2 - E_y^2\\E_{45}^2 - E_{-45}^2\\E_R^2 - E_L^2\end{bmatrix}
\end{equation}

where $E_x$, $E_y$, $E_{45}$, and $E_{-45}$ are projections onto horizontal, vertical, 45-degree, -45-degree linear polarizations, respectively and $E_R$ and $E_L$ are projections onto right- and left-circular polarizations, respectively.  To calibrate the device, we input each of these individual polarization components and observe the transmission into each of the four quadrants.  Then, we form:
\begin{equation}
\begin{aligned}
    \underline{\underline{M}} \underline{\underline{\sigma}} = \underline{\underline{\tau}}\\
    \underline{\underline{\sigma}} = \begin{bmatrix}
    \vec{S_x}&\vec{S_y}&\vec{S}_{45}&\vec{S}_{-45}&\vec{S_R}&\vec{S_L}
    \end{bmatrix} \in \mathbb{R}^{4x6}  \\
    \underline{\underline{\tau}} = \begin{bmatrix}
    \vec{T_x}&\vec{T_y}&\vec{T}_{45}&\vec{T}_{-45}&\vec{T_R}&\vec{T_L}
    \end{bmatrix} \in \mathbb{R}^{4x6}\\
    \underline{\underline{M}} \in \mathbb{R}^{4x4}
\end{aligned}
\end{equation}
where $\vec{S_x} = \begin{bmatrix}1 &1& 0& 0\end{bmatrix}^\dagger$, $\vec{S_y} = \begin{bmatrix}1 &-1& 0& 0\end{bmatrix}^\dagger$, $\vec{S}_{45} = \begin{bmatrix}1 &0& 1& 0\end{bmatrix}^\dagger$, $\vec{S}_{-45} = \begin{bmatrix}1 &0& -1& 0\end{bmatrix}^\dagger$, $\vec{S_R} = \begin{bmatrix}1 &0& 0& 1\end{bmatrix}^\dagger$, $\vec{S_L} = \begin{bmatrix}1 &0& 0& -1\end{bmatrix}^\dagger$ and $\vec{T_{\alpha}}$ are the four quadrant transmissions under excitation by the the $\vec{S_{\alpha}}$ state.  We solve for $\underline{\underline{M}}$ by taking the pseudo-inverse of $\underline{\underline{\sigma}}$ and applying it on the right side, $\underline{\underline{M}}=\underline{\underline{\tau}}\underline{\underline{\sigma}}^\dagger$. Then, we form the solution or reconstruction matrix by taking the inverse of $\underline{\underline{M}}$, such that given a set of measurements $\vec{T}$, we compute the Stokes parameters as $\vec{S} = \underline{\underline{M}}^{-1}\vec{T}$.  We note this calibration could alternatively be done with the four analyzer states used in the design and we expect the results would be similar.

\subsubsection*{\textbf{Reconstructing Pure Polarization States}}

The reconstruction method applied to pure polarization states is shown in \verb+Fig. S9+ for different amounts of added noise in the transmission measurements to simulate different signal-to-noise ratios in the sensor detection.  For $p$ added noise, we add a normally distributed random variable with a mean of $0$ and a standard deviation equal to $p*T_{avg}$ where $T_{avg}$ is the mean transmission across the four quadrant transmissions.  As can be seen for increasing noise, the $S_3$ parameter is the most susceptible to a reduced signal-to-noise ratio.  This is likely due to the circular polarization analyzer state exhibiting the lowest contrast and the $S_3$ Stokes parameter being a direct measure of the handedness of the circular polarization in the input.

\subsubsection*{\textbf{Reconstructing Mixed Polarization States}}

The use of four projective measurements means information about partially polarized input states is contained in the quadrant transmissions.  To test our ability to recover this property, we consider the situation where the polarization vector input into the device is randomly changing.  We input a series of random polarization states into the device, and average the resulting quadrant transmission values for each quadrant.  From these averaged transmission values, we reconstruct the Stokes vector in the same way as above.  This reconstructed vector is compared to the averaged Stokes vectors for all the states input into the device.  The degree of polarization of the light is computed as $p=\frac{\sqrt{S_1^2 + S_2^2+S_3^2}}{S_0}$.

\verb+Fig. S10+ shows the results of reconstructing mixed polarization states.  As the number of averaged states increases, the degree of polarization starts dropping.  When noise is added per averaged state (using the same type of distribution as above), the squared error for the reconstruction is highest for the smaller number of averaged states.  As this number of states increases, the fluctuating noise term starts averaging to zero thus decreasing the overall effect of noise on the reconstruction.

\subsection*{Angular Momentum Sorting Device Outside of Design Points}

\verb+Fig. S11+ and \verb+Fig. S12+ demonstrate the behavior of the angular momentum sorting device for different values of spin and OAM, respectively, than the design states.  In an optical communication application, controlling the behavior of the device at these alternate points will depend on the amount noise present and mode distortion between communication links.  However, in an advanced imaging context where information about the scene is inferred through the spatially resolved projection of the input onto different angular momentum states, the response of the device to other mode inputs needs to be at least characterized if not explicitly designed for the given application.  As a note, the optimization technique used here was not directed to explicitly minimize or control the behavior of the device under these other excitations.  By adding more simulations to each iteration to capture the effect of illuminating with these other modes, we can compute a gradient that either enables control over the quadrant these other modes couple to or extinguishes their transmission.

\subsubsection*{\textbf{Illumination with Different Spin Values}}

In \verb+Fig. S11+, we observe the device behaves similarly upon a flip in the handedness of the circular polarization for each angular momentum state.  This can be seen through similar contrast and transmission profiles albeit at lower overall values.  Thus, the optimization solution for the device relied primarily on the different OAM values for splitting and does not have strong polarization discriminating behavior.

\subsubsection*{\textbf{Illumination with Different Spin Values}}

In \verb+Fig. S12+, we observe the device output changes drastically when illuminated with different OAM values.  Most of the light for each of the four states goes to the quadrants designed for the original higher design OAM values (i.e. - $l=-2,+2$).  This is the reason for the negative contrast in the other two quadrants.  Further, overall transmission values are significantly reduced with the higher transmission occurring for OAM values closer to the design points (i.e - $l=-3,+3$).

\newpage
\section*{Supplementary Figures}

\setcounter{figure}{0}

\renewcommand{\figurename}{Fig.}
\renewcommand{\thefigure}{S\arabic{figure}}
\begin{figure}[h]
\thisfloatpagestyle{plain}
\begin{center}
    \includegraphics{"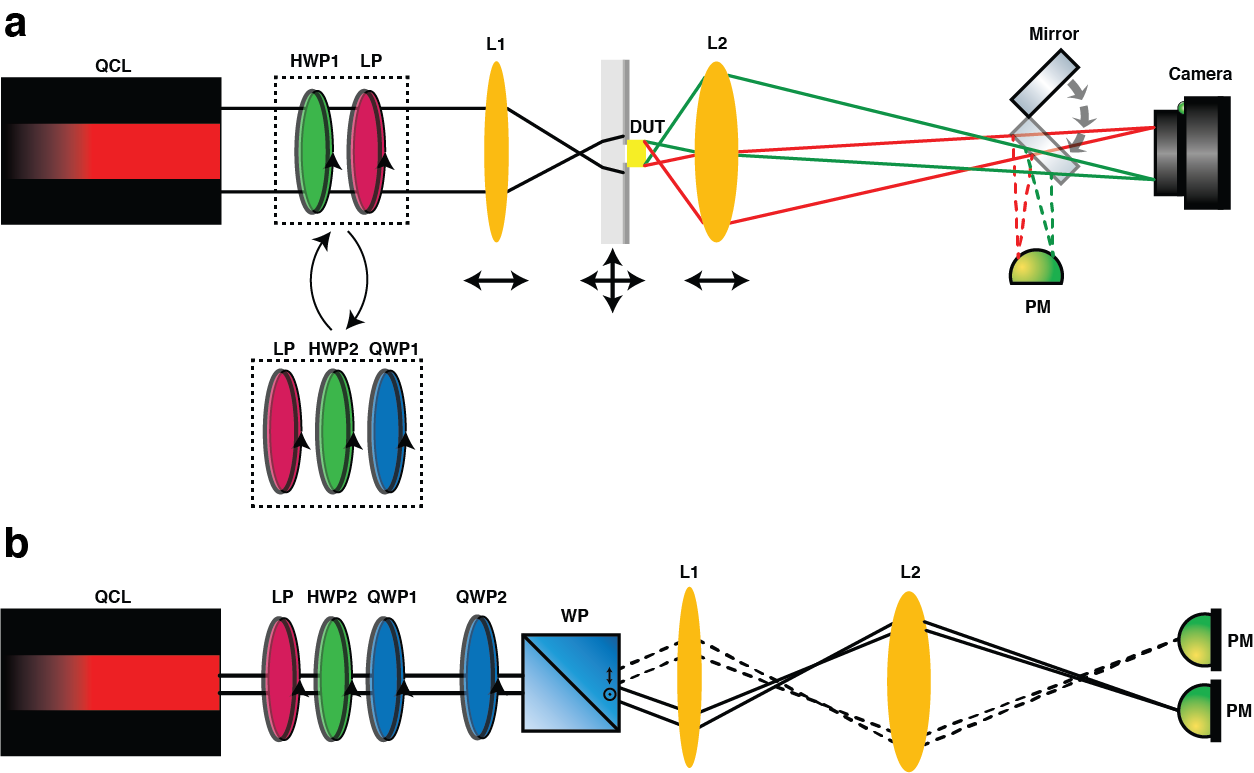"}
\end{center}
\caption{Optical setup for characterization of multispectral and polarimetry devices.  \textbf{(a)} Configuration for imaging of device focal plane and power normalization.  Without the mirror in place, the lens images focal planes of the device onto the camera.  Normalization of the device transmission is done with the mirror and power meter path of the setup.  For these measurements, net power through an empty aperture is used to normalize net power through an aperture of the same size with the device on top of it.  The power meter is aligned to the beam center, which is aligned to the pinhole centers during measurement.  QCL: MIRcat-QT Mid-IR Quantum Cascade Laser (DRS Daylight Solutions); HWP1: Thorlabs WPLH05M-4500, Low-Order 4.5$\mu m$ Half-Wave Plate; HWP2: Thorlabs WPLH05M-5300, Zero-Order 5.3$\mu m$ Half-Wave Plate; QWP 1: Thorlabs WPLQ05M-4500, Low-Order 4.5$\mu m$ Quarter-Wave Plate; LP: Thorlabs WP25M-IRA, Wire Grid Polarizer; L1: Thorlabs AL72525-E1, ZnSe Aspheric Lens, NA=0.42; L2: Thorlabs AL72512-E1, ZnSe Aspheric Lens, NA=0.67; Camera: Electrophysics PV320L IR Camera.  \textbf{(b)} Configuration for verifying the polarization states used to test the Stokes polarimetry device.  The second quarter wave plate is moved to three distinct positions and the power in each linear polarization component separated by the Wollaston prism is recorded. QWP2: Thorlabs WPLQ05M-3500, Low-Order 3.5$\mu m$ Quarter-Wave Plate; WP: Thorlabs WPM10, Wollaston Prism.}
\end{figure}\label{figS1}

\renewcommand{\figurename}{Fig.}
\renewcommand{\thefigure}{S\arabic{figure}}
\begin{figure}[h]
\thisfloatpagestyle{plain}
\begin{center}
    \includegraphics{"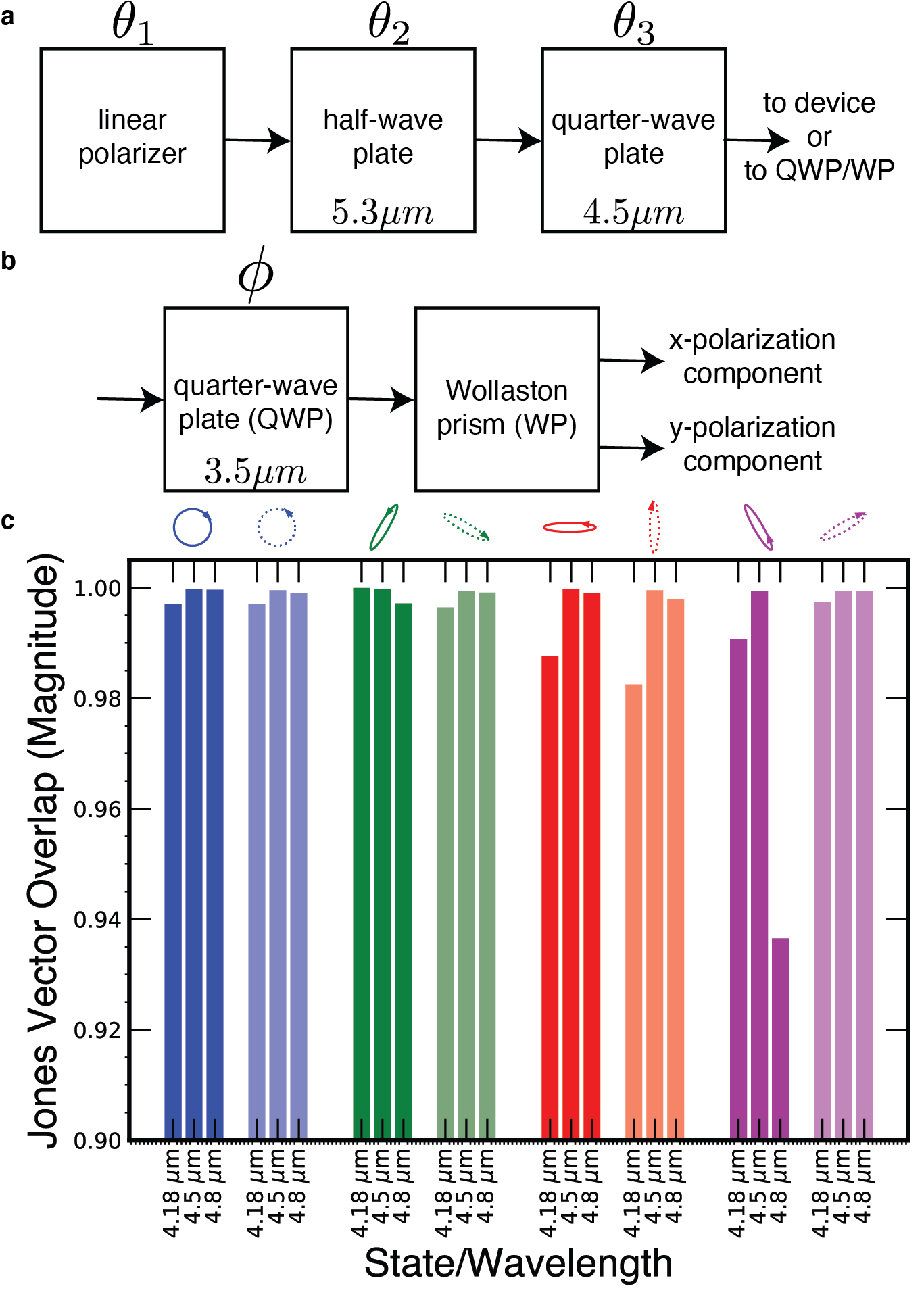"}
\end{center}
\caption{Stokes state creation and verification.  \textbf{(a)} Polarization states are created through choice of angles of the linear polarizer, half-wave plate, and quarter-wave plate ($\theta_1$, $\theta_2$, and $\theta_3$).  \textbf{(b)} Each state is verified by measuring the horizontal and vertical polarization component magnitudes output from the Wollaston prism after the state passes through a quarter-wave plate under three different rotations, $\phi$.  \textbf{(c)} Plot of measured Jones vector overlap for the $4$ analyzer states and their $4$ orthogonal complements for each measurement wavelength used in the experiment.}
\end{figure}\label{figS2}

\renewcommand{\figurename}{Fig.}
\renewcommand{\thefigure}{S\arabic{figure}}
\begin{figure}[h]
\thisfloatpagestyle{plain}
\begin{center}
    \includegraphics{"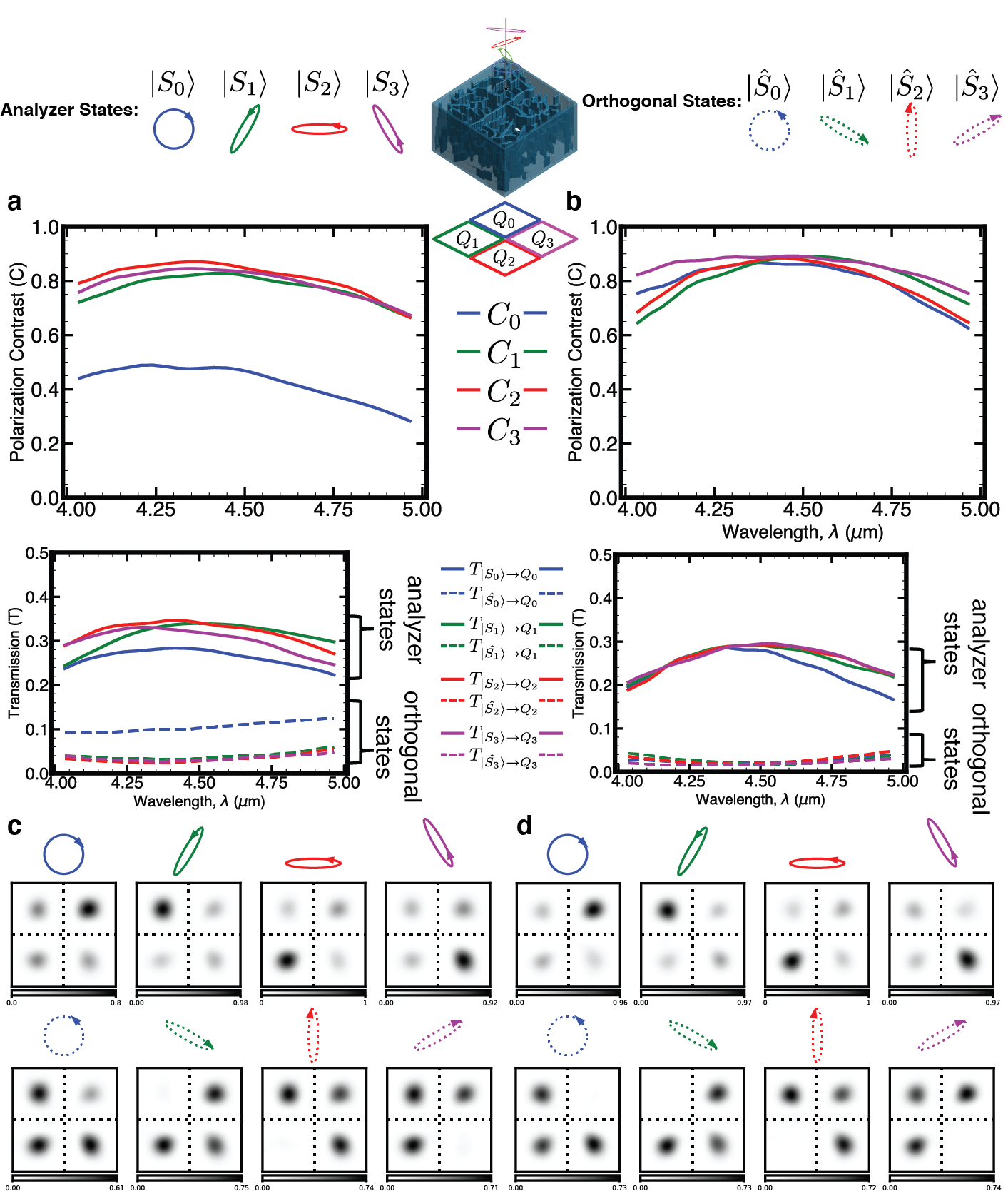"}
\end{center}
\caption{Simulation performance for Stokes polarimetry device with additional degrees of freedom compared to Stokes polarimetry device from the main text.  \textbf{(a)} Polarization contrast ($C)$) and transmission ($T$) for device from the main text showing low contrast for the circular polarization analyzer state.  For input $k$, $C_k=\frac{T_{\ket{S_k} \rightarrow Q_k } - T_{\ket{\hat{S_k}} \rightarrow Q_k}}{T_{\ket{S_k} \rightarrow Q_k } + T_{\ket{\hat{S_k}} \rightarrow Q_k}}$.  \textbf{(b)} Polarization contrast and transmission for device with additional degrees of freedom showing high contrast for all four analyzer states at the cost of slightly reduced analyzer state transmission. \textbf{(c)} Focal intensity images for device from the main text with the top row showing the analyzer states and the bottom row showing their orthogonal complements.  Intensity units are arbitrary but comparable between all plots in (c). The focal plane size is same as device aperture ($30\mu m$ x ($30 \mu m$). \textbf{(d)} Focal intensity image comparison for the device with additional degrees of freedom. Intensity units are arbitrary but comparable between all plots in (d). The focal plane size is the same as device aperture ($30\mu m$ x ($30 \mu m$).}
\end{figure}\label{figS3}

\renewcommand{\figurename}{Fig.}
\renewcommand{\thefigure}{S\arabic{figure}}
\begin{figure}[h]
\thisfloatpagestyle{plain}
\begin{center}
    \includegraphics{"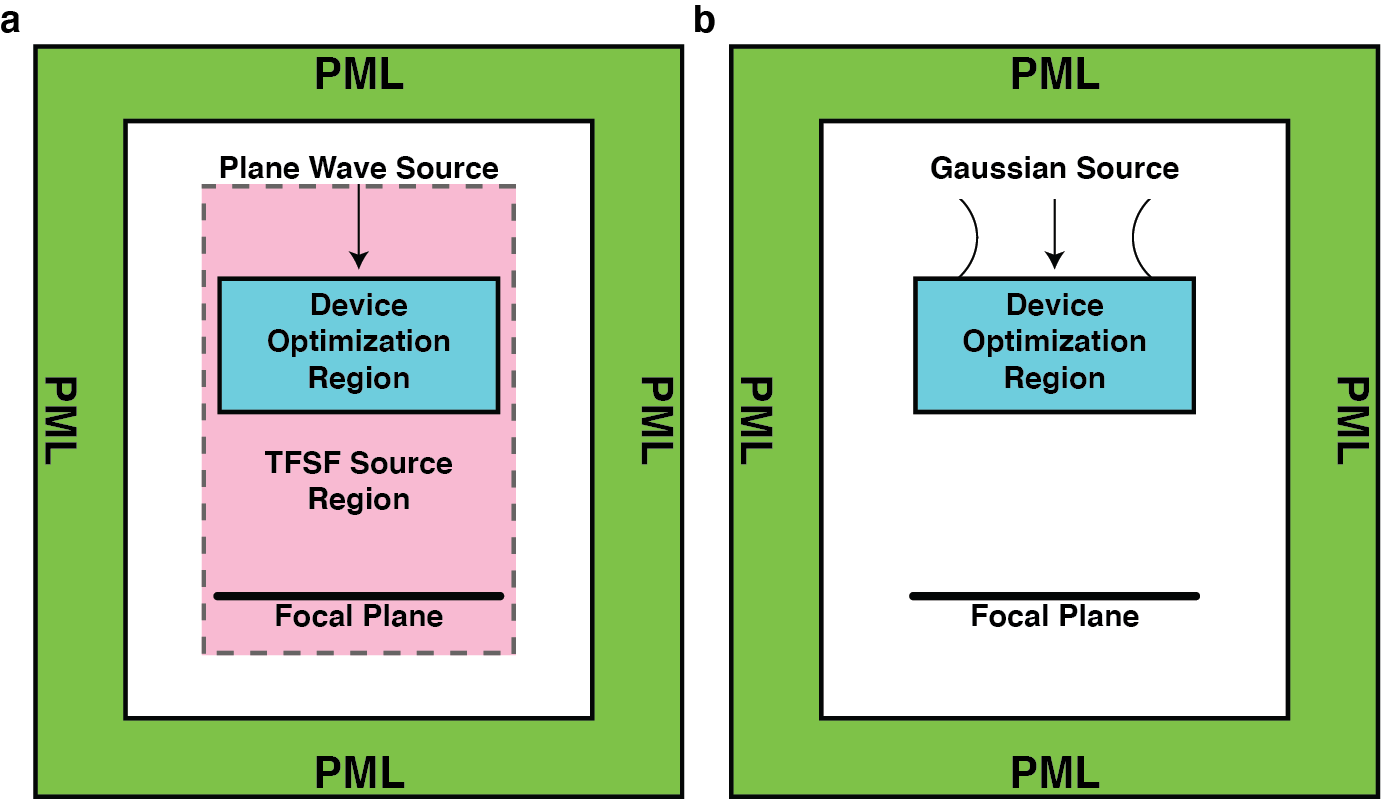"}
\end{center}
\caption{Schematic of simulation geometry for optimization and evaluation. \textbf{(a)} Simulation geometry for optimization of the multispectral and Stokes polarimetry devices using a plane wave excitation.  The angular momentum devices are optimized using focused angular momentum states with different  circular polarization handedness for spin.  \textbf{(b)} Evaluation geometry for the multispectral and Stokes polarimetry devices where the plane wave excitation is replaced with a defocused Gaussian source intended to match with the experimental source.  Angular momentum devices are evaluated with the same sources as used for optimization.}
\end{figure}\label{figS4}

\renewcommand{\figurename}{Fig.}
\renewcommand{\thefigure}{S\arabic{figure}}
\begin{figure}[h]
\thisfloatpagestyle{plain}
\begin{center}
    \includegraphics{"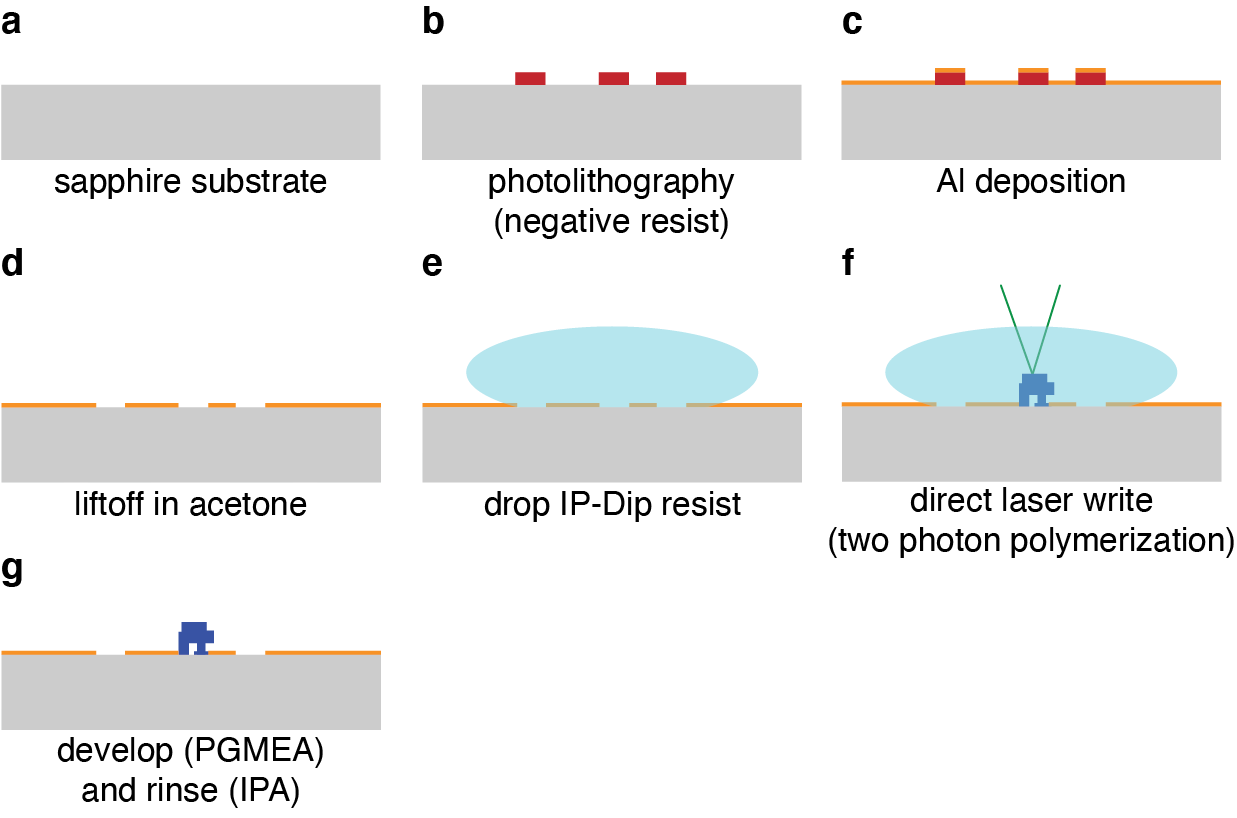"}
\end{center}
\caption{Schematic of fabrication process. \textbf{(a)} Fabrication starts with a sapphire substrate (${\text{Al}_2\text{O}_3}$, C-plane (0001), double side polished, 2-inch diameter, 0.5mm thickness). \textbf{(b)} Using a negative tone photoresist, apertures are patterned onto the substrate using photolithography. \textbf{(c)} After direct oxygen and argon plasma cleaning to remove undesired residual resist on the substrate, 150${nm}$ of Al is deposited on top using an electron beam evaporator. \textbf{(d)} The liftoff procedure is finished in acetone to remove remaining photoresist followed by cleaning in IPA and then DI water. \textbf{(e)} The IP-Dip resist from Nanoscribe is dropped onto the substrate. \textbf{(f)} Alignment is done by keeping the laser power below polymerization threshold and using fluorescence from its focused spot to align to the aperture centers for printing. \textbf{(g)} Development in propylene glycol methyl ether acetate (PGMEA) for 20 minutes followed by two three-minute rinses in IPA reveals the final device.}
\end{figure}\label{figS5}

\renewcommand{\figurename}{Fig.}
\renewcommand{\thefigure}{S\arabic{figure}}
\begin{figure}
\thisfloatpagestyle{plain}
\begin{center}
    \includegraphics{"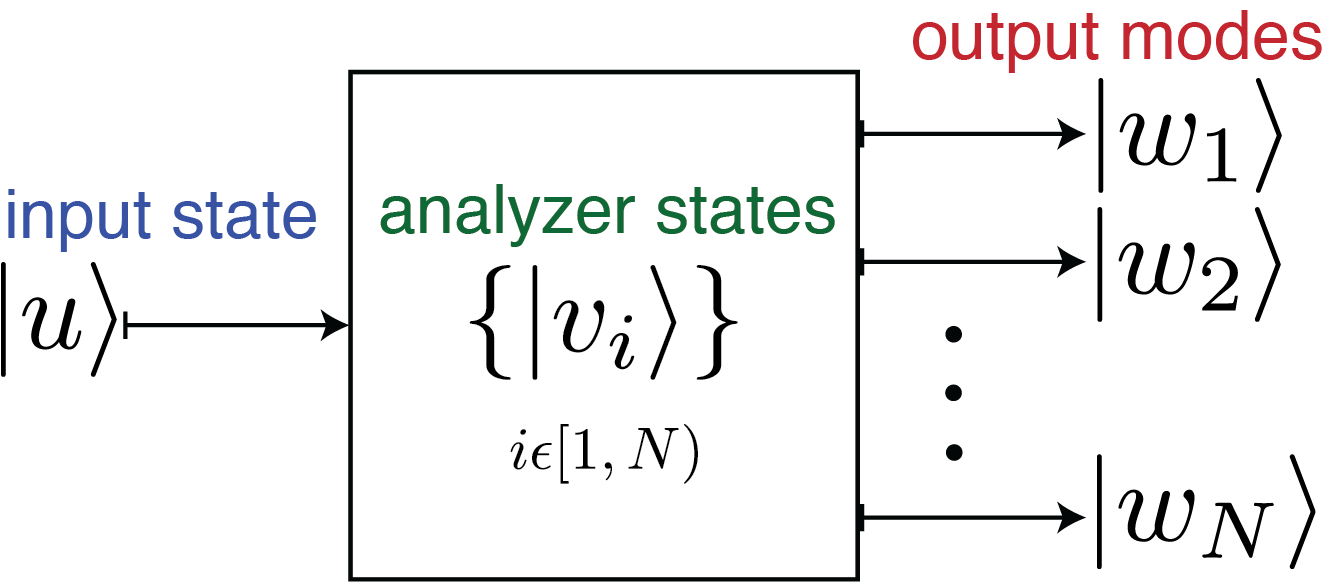"}
\end{center}
\caption{Conceptual diagram of the Stokes polarimetry device.  The device acts to project an input state onto four outgoing states depending on its overlap with each analyzer state.}
\end{figure}\label{figS6}

\renewcommand{\figurename}{Fig.}
\renewcommand{\thefigure}{S\arabic{figure}}
\begin{figure}[h]
\thisfloatpagestyle{plain}
\begin{center}
    \includegraphics{"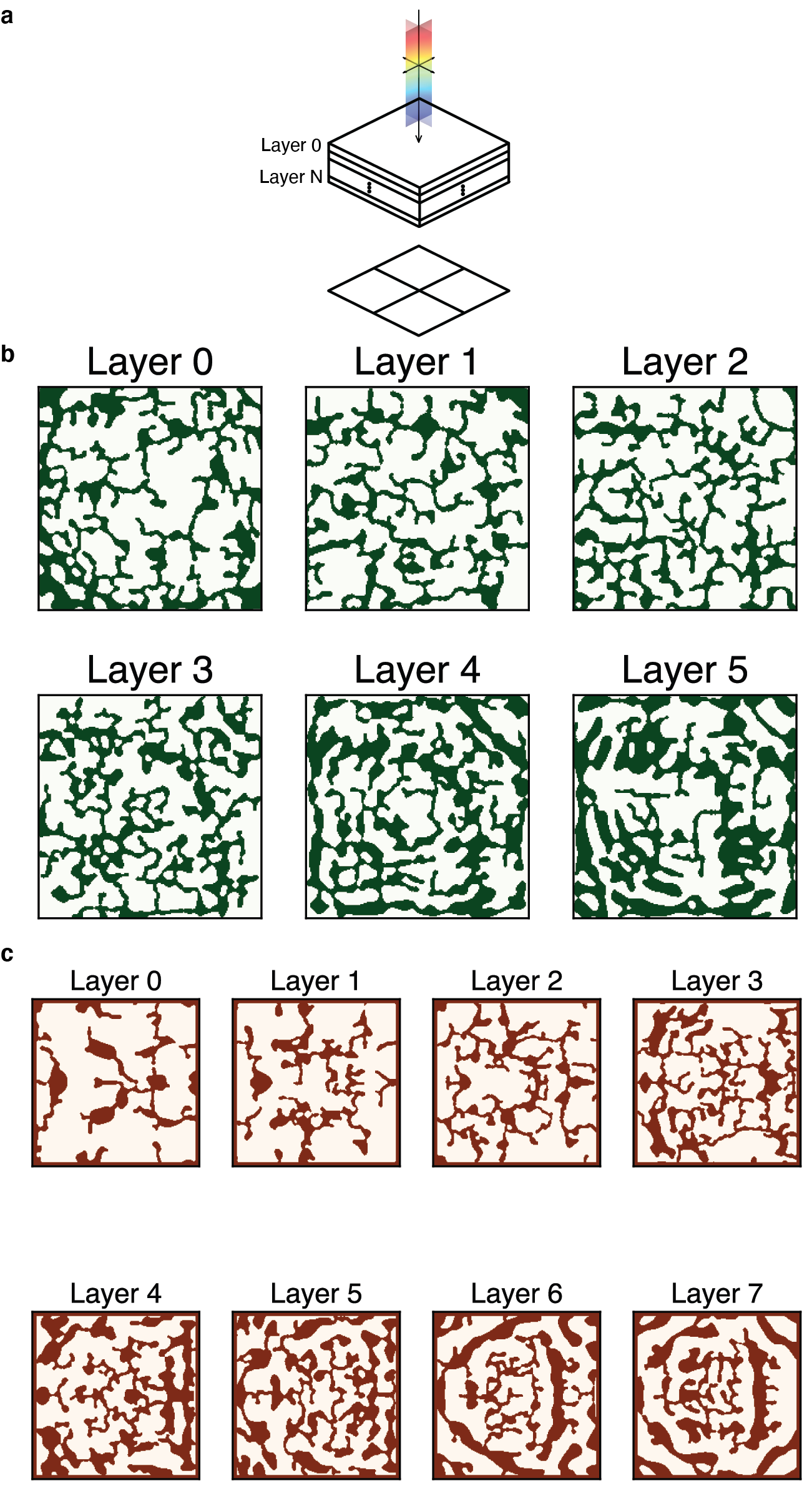"}
\end{center}

\caption{Index of refraction profiles for multispectral (b) and angular momentum (c) devices.  Dark colors are IP-Dip polymer and light areas are void.  \textbf{()} Schematic of geometry showing the location of each labeled layer. Each layer is $3 \mu m$ thick for the multispectral device and $2.4\mu m$ thick for the angular momentum device.  \textbf{(b)} Multispectral and linear polarization device index profile with each layer $30.15 \mu m$ x $30.15 \mu m$.  \textbf{(d)} Angular momentum device index profile with each layer $30.15 \mu m$ x $30.15 \mu m$.}
\end{figure}\label{figS7}

\renewcommand{\figurename}{Fig.}
\renewcommand{\thefigure}{S\arabic{figure}}
\begin{figure}[h]
\thisfloatpagestyle{plain}
\begin{center}
    \includegraphics{"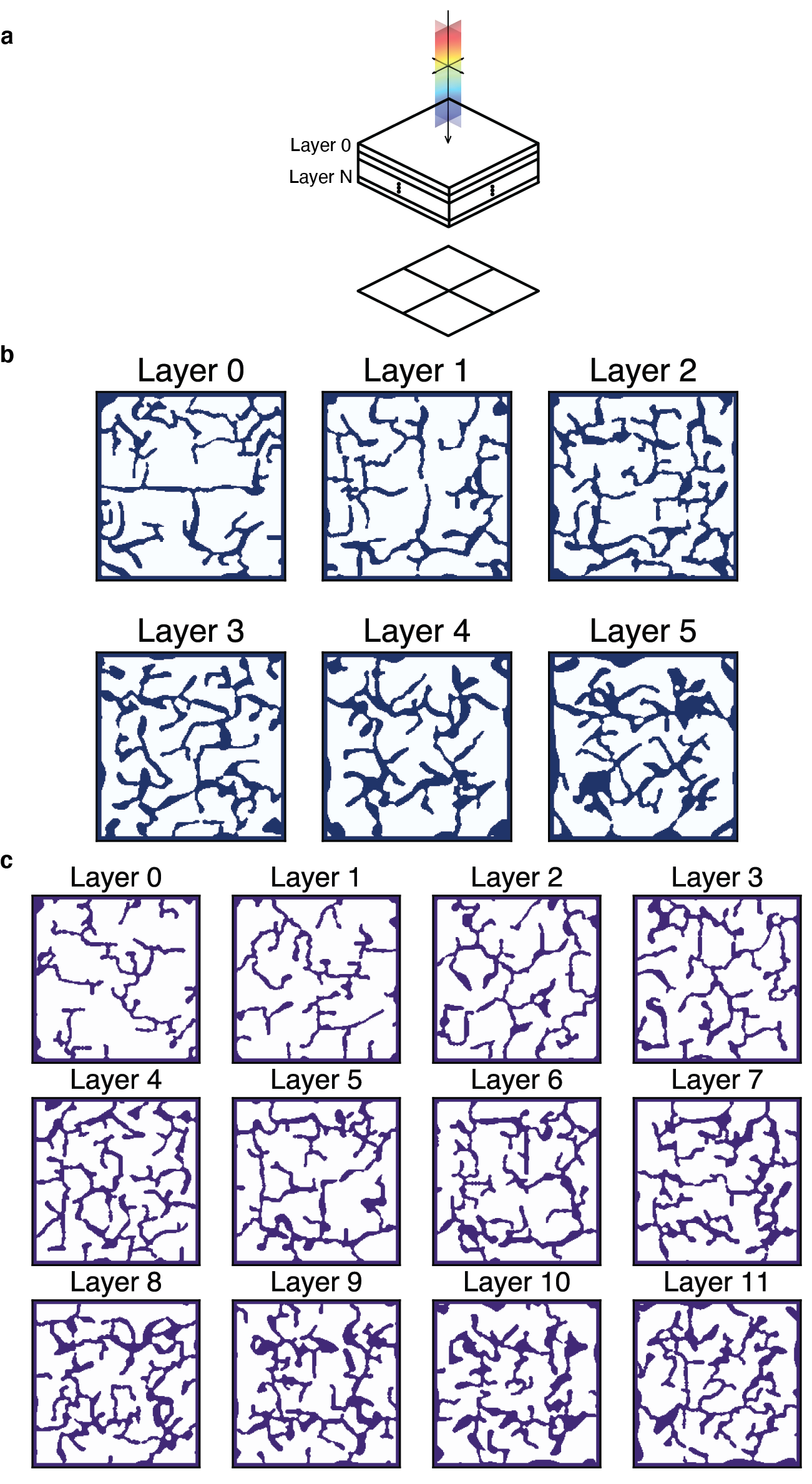"}
\end{center}
\caption{Index of refraction profiles for Stokes polarimetry device from main text (b) and Stokes polarimetry device with more degrees of freedom from supplement (c).  Dark colors are IP-Dip polymer and light areas are void.  \textbf{(a)} Schematic of geometry showing location of each labeled layer. Each layer is $3 \mu m$ thick. \textbf{(b)} Stokes polarimetry device from main text index profile with each layer $30 \mu m$ x $30 \mu m$. textbf{(d)} Stokes polarimetry device with additional degrees of freedom index profile with each layer $30 \mu m$ x $30 \mu m$.}
\end{figure}\label{figS8}

\renewcommand{\figurename}{Fig.}
\renewcommand{\thefigure}{S\arabic{figure}}
\begin{figure}[h]
\thisfloatpagestyle{plain}
\begin{center}
    \includegraphics{"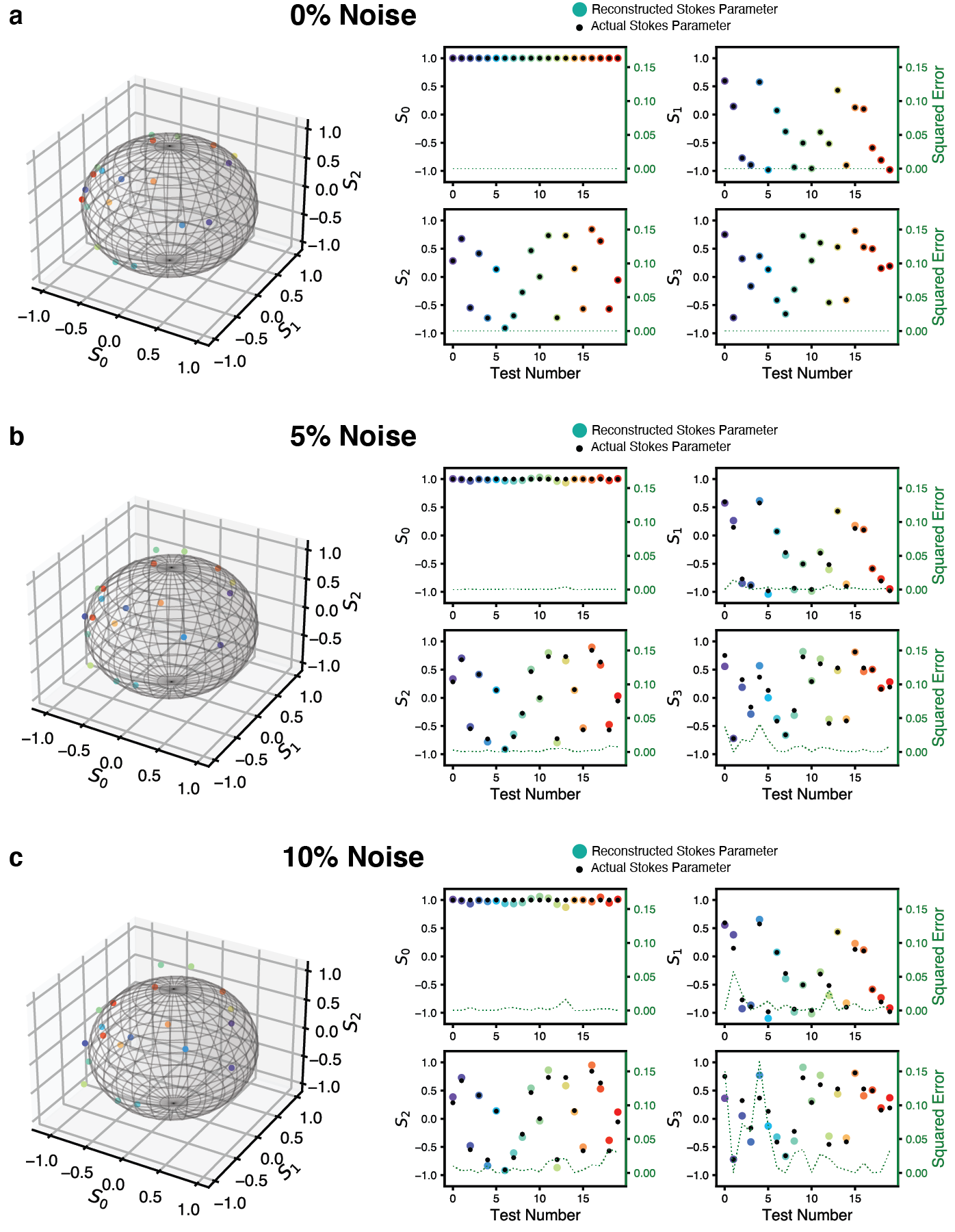"}
\end{center}
\caption{Stokes polarimetry reconstruction in simulation using the device from the main text under random pure polarization inputs at a wavelength of $4.5\mu m$.  \textbf{(a)} Reconstructed state locations shown on Poincaré sphere on the left and comparison of the reconstructed Stokes parameters to the actual ones shown on the right with the associated squared error (green dashed line, right y-axis). \textbf{(b)} Same plots as (a) but with $5\%$ added noise. \textbf{(c)} Same plots as (a) but with $10\%$ added noise.}
\end{figure}\label{figS9}

\renewcommand{\figurename}{Fig.}
\renewcommand{\thefigure}{S\arabic{figure}}
\begin{figure}[h]
\thisfloatpagestyle{plain}
\begin{center}
    \includegraphics{"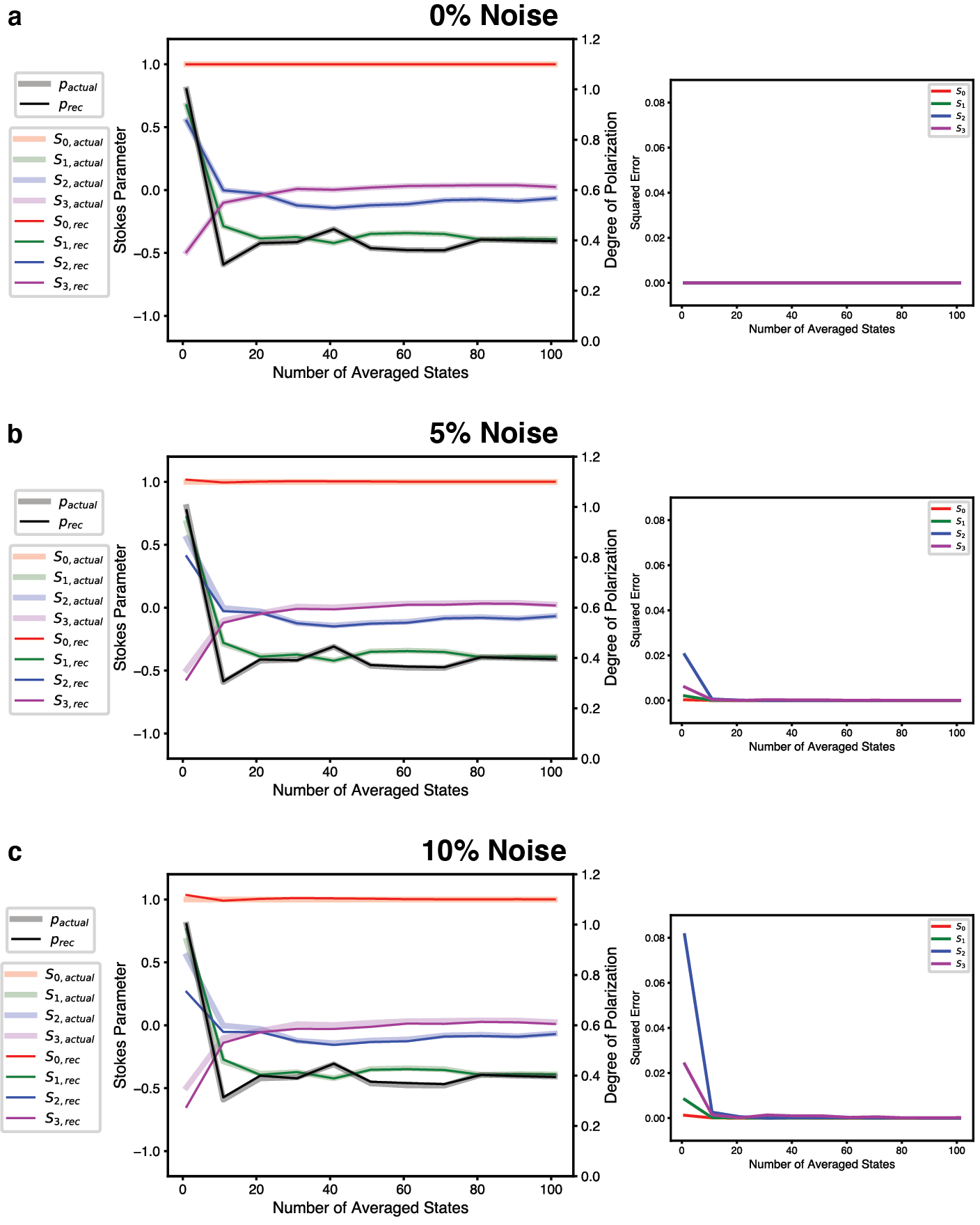"}
\end{center}
\caption{Stokes polarimetry reconstruction in simulation using the device from the main text under random mixed polarization inputs at a wavelength of $4.5\mu m$.  \textbf{(a)} Reconstructed state locations broken down by Stokes parameter as well as degree of polarization compared to actual ones shown on the left with the associated squared error per Stokes parameter shown on the right. \textbf{(b)} Same plots as (a) but with $5\%$ added noise. \textbf{(c)} Same plots as (a) but with $10\%$ added noise.}
\end{figure}\label{figS10}




\renewcommand{\figurename}{Fig.}
\renewcommand{\thefigure}{S\arabic{figure}}
\begin{figure}[h]
\thisfloatpagestyle{plain}
\begin{center}
    \includegraphics{"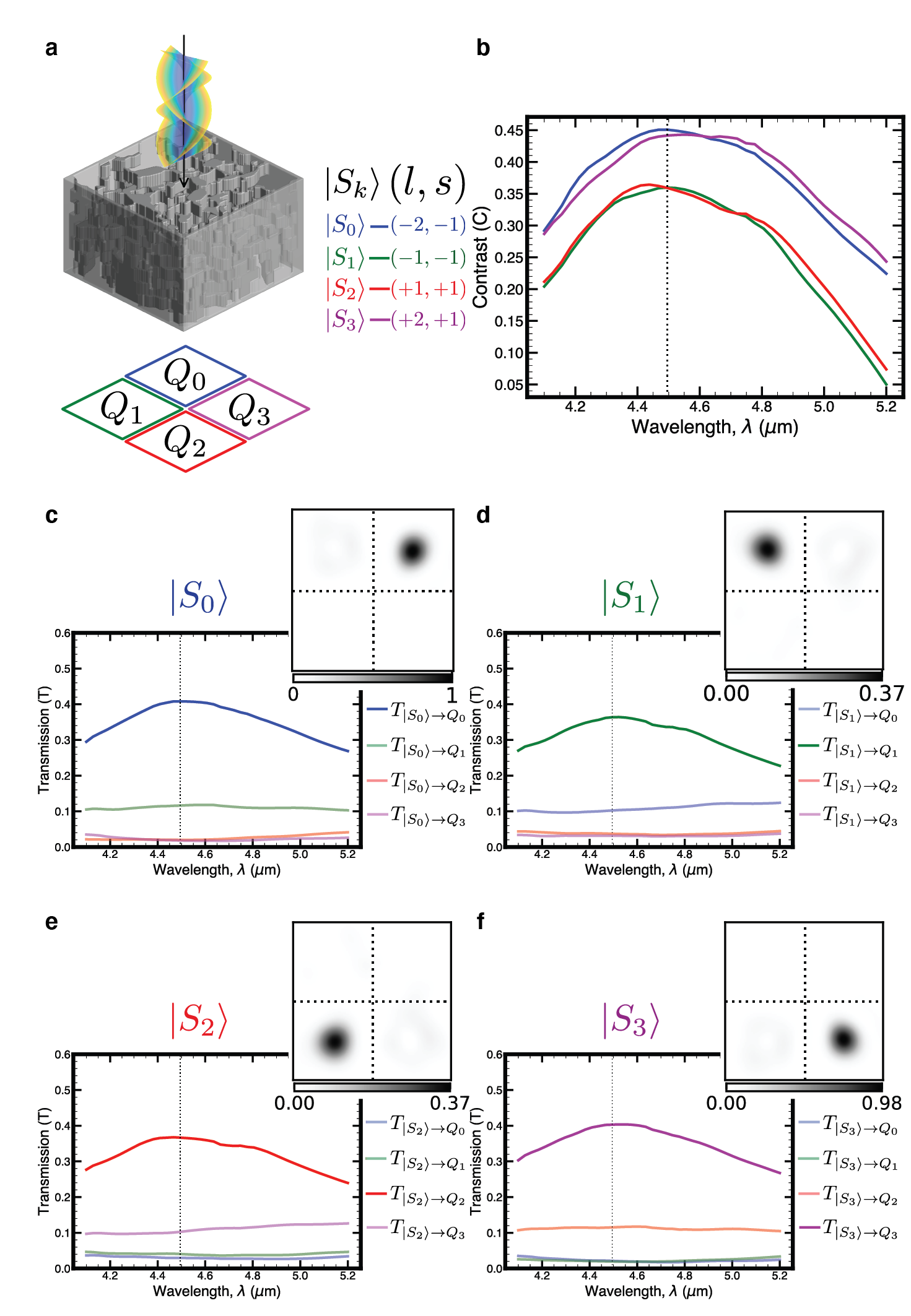"}
\end{center}

\caption{Simulation of angular momentum sorting device for the same OAM, but different spin values than the design states. \textbf{(a)} Schematic of device and focal plane quadrants. \textbf{(b)} Contrast for sorting each state ($C\in[-1, 1]$) defined by the transmission of a state into the desired quadrant versus the transmission into the rest of the focal plane.  For source $k$, $C_k = \frac{T_{\ket{S_k} \rightarrow Q_k } - \sum_{i \neq k }{ T_{\ket{S_k} \rightarrow Q_i}}}{T_{\ket{S_k} \rightarrow Q_k } + \sum_{i \neq k }{ T_{\ket{S_k} \rightarrow Q_i}}}$. \textbf{(c)} Transmission spectrum for ($l=-2$, $s=-1$) input with desired quadrant transmission in blue.  Transmission is normalized by power through the device aperture with no device present.  Inset: Intensity at focal plane (arbitrary units, but comparable to other intensity plots in figure). \textbf{(d-f)} Same plots as (c), but for ($l=-1$, $s=-1$), ($l=+1$, $s=+1$), ($l=+2$, $s=+1$), respectively.}
\end{figure}\label{figS11}

\renewcommand{\figurename}{Fig.}
\renewcommand{\thefigure}{S\arabic{figure}}
\begin{figure}[h]
\thisfloatpagestyle{plain}
\begin{center}
    \includegraphics{"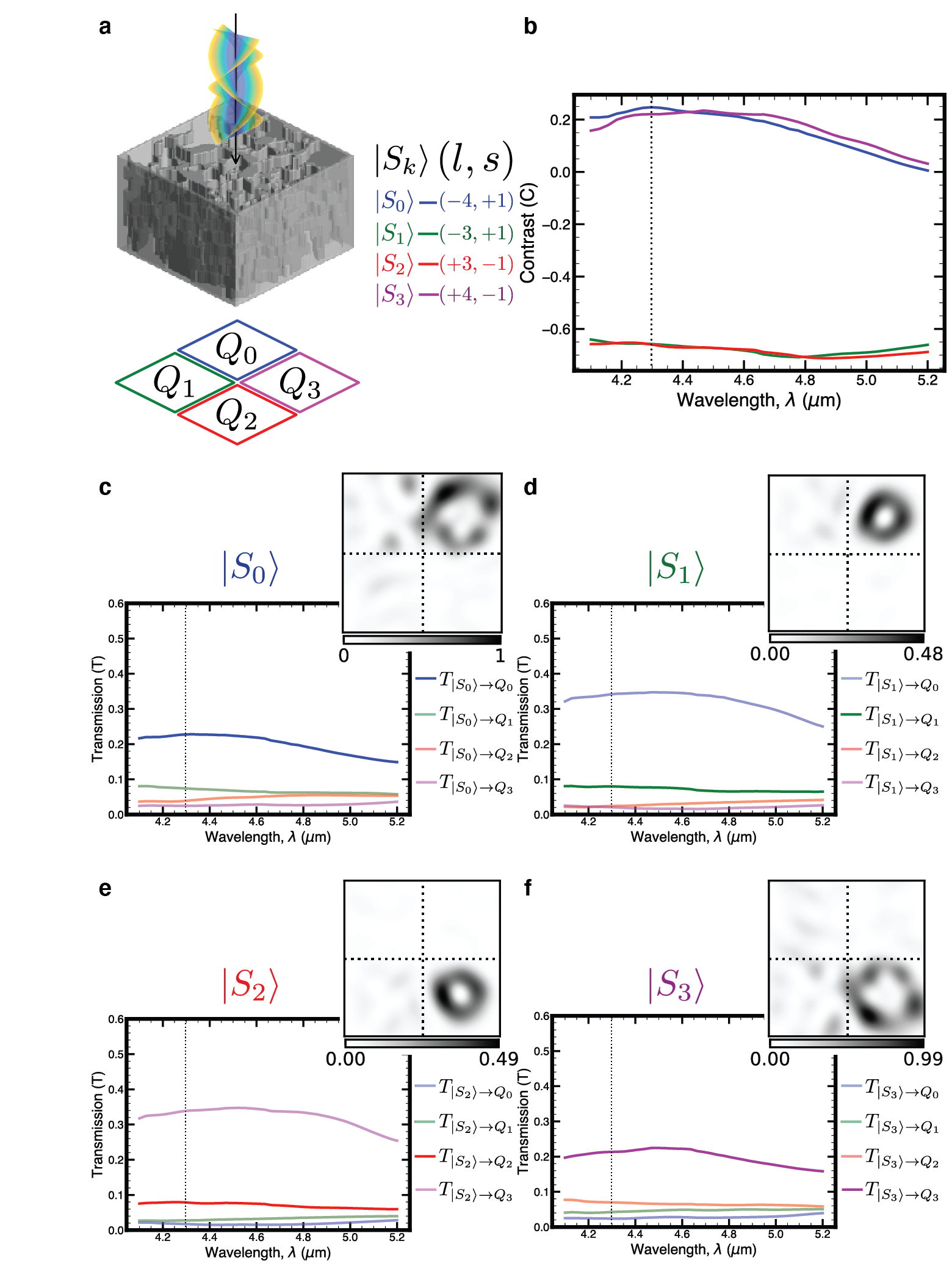"}
\end{center}

\caption{Simulation of angular momentum sorting device for the same spin values, but different OAM than the design points. \textbf{(a)} Schematic of device and focal plane quadrants. \textbf{(b)} Contrast for sorting each state ($C\in[-1, 1]$) defined by the transmission of a state into the desired quadrant versus the transmission into the rest of the focal plane.  For source $k$, $C_k = \frac{T_{\ket{S_k} \rightarrow Q_k } - \sum_{i \neq k }{ T_{\ket{S_k} \rightarrow Q_i}}}{T_{\ket{S_k} \rightarrow Q_k } + \sum_{i \neq k }{ T_{\ket{S_k} \rightarrow Q_i}}}$. \textbf{(c)} Transmission spectrum for ($l=-4$, $s=+1$) input with desired quadrant transmission in blue.  Transmission is normalized by power through the device aperture with no device present.  Inset: Intensity at focal plane (arbitrary units, but comparable to other intensity plots in figure). \textbf{(d-f)} Same plots as (c), but for ($l=-3$, $s=+1$), ($l=+3$, $s=-1$), ($l=+4$, $s=-1$), respectively.}
\end{figure}\label{figS12}

\end{document}